\documentclass{article}

\usepackage{PRIMEarxiv}

\usepackage{abstract}
\usepackage[utf8]{inputenc}
\usepackage[T1]{fontenc}
\usepackage{hyperref}
\usepackage{url}
\usepackage{booktabs}
\usepackage{array}
\usepackage{amsfonts}
\usepackage{amssymb}
\usepackage{amsmath}
\usepackage{amsthm}
\usepackage{nicefrac}
\usepackage{microtype}
\usepackage{fancyhdr}
\usepackage{graphicx}
\usepackage[table]{xcolor}
\usepackage{placeins}
\usepackage{quantikz}
\usepackage{tikz}

\graphicspath{{media/}}
\hypersetup{hidelinks}

\newtheorem{protocol}{Protocol}
\newtheorem{primitive}{Primitive}
\newtheorem{proposition}{Proposition}[section]
\newtheorem{lemma}{Lemma}[section]
\newtheorem{remark}{Remark}[section]
\newtheorem{theorem}{Theorem}
\newtheorem{criterion}{Criterion}
\newtheorem{assumption}{Assumption}

\newcommand{\SWAP}{\operatorname{SWAP}}
\newcommand{\CNOT}{\operatorname{CNOT}}
\newcommand{\CZ}{\operatorname{CZ}}
\newcommand{\CCZ}{\operatorname{CCZ}}
\newcommand{\ApplyPrivateCCZ}{\textsc{ApplyPrivateCCZ}}

\newcolumntype{L}[1]{>{\raggedright\arraybackslash}p{#1}}

\title{Private Delegated Quantum Computing for User-Level and Industry-Level Settings}

\author{%
 Alejandro Mata Ali \\
 Instituto Tecnol\'ogico de Castilla y Le\'on, Burgos, Spain\\
 Quantum Information and Quantum Computing Group, QuantumQuipu,\\
 National University of San Marcos, Av.\ Germán Amézaga s/n.\ Ciudad Universitaria, Lima, Perú \\
 \texttt{alejandro.mata.ali@gmail.com} \\
 \AND
 Adriano Mauricio Lusso \\
 Quantum Information and Quantum Computing Group, QuantumQuipu,\\
 National University of San Marcos, Av.\ Germán Amézaga s/n.\ Ciudad Universitaria, Lima, Perú \\
 Universidad Nacional del Comahue, (8300) Neuquén, Argentina \\
 \texttt{lussoadriano@gmail.com} \\
 \AND
 Edgar Mencia \\
 Quantum Information and Quantum Computing Group, QuantumQuipu,\\
 National University of San Marcos, Av.\ Germán Amézaga s/n.\ Ciudad Universitaria, Lima, Perú \\
 Thinkcomm SRL, (1527) Asunción, Paraguay \\
 \texttt{edmenciab@gmail.com} \\
}

\begin{document}
\maketitle

\begin{abstract}
 We present a modular hierarchy of private delegated quantum computation protocols tailored to user-level and industry-level settings and parameterized by the quantum resources available to the client. For each protocol, we specify the client capabilities, delegated gate set, adversarial model, transcript leakage and resulting privacy claims. The hierarchy separates QOTP state privacy under declared leakage from leakage-dependent transcript-level angle ambiguity, compiler- and leakage-function-dependent structural privacy, and output privacy, clarifies when public Clifford operations can be evaluated on quantum-one-time-pad encrypted data by classical key updates, and identifies where non-Clifford privacy, non-collusion or additional primitives are required. The classical-client branch uses a persistent common-node, matching-hidden split-QOTP together with shuffled finite-grid $r$-share sign-randomized angle sharing to obtain leakage-relative state hiding under an explicit \(\epsilon_{\mathrm{key}}\) key-hiding condition and transcript-level unlinkability under hidden-matching assumptions under an explicit non-total-collusion and leakage model. The angle-sharing primitives provide transcript ambiguity under explicit leakage assumptions, not universal blindness. The trap-based layer provides detection under stated assumptions, but it is not a stand-alone malicious-security proof.
\end{abstract}

\tableofcontents

\section{Introduction}
Quantum computing has evolved from a largely theoretical concept into an active field of research and investment, driven by its potential to provide computational advantages in a wide range of scientific and industrial applications. However, access to quantum hardware remains a major bottleneck for both individual users and companies, since even small-scale quantum devices are expensive to build, operate, and maintain.

A natural way to address this bottleneck is delegated quantum computation. In this model, a client with little or no quantum computational resources delegates the required quantum operations to a server with the necessary capabilities. This is a quantum analogue of the use of classical cloud servers.

However, in several industrial cases it is essential to preserve the privacy of the computation's input data, output data, and even its description. A clear example is a company attempting to solve a combinatorial optimization problem, such as the Quadratic knapsack problem, using a Quadratic Unconstrained Binary Optimization (QUBO) formulation \cite{QUBO} adapted to the characteristics of its problem and solved using the Quantum Approximate Optimization Algorithm (QAOA) \cite{QAOA}. In this case, the values of the coefficients would reveal private information about the problem to the server, potentially related to an economic, telecommunications or transportation project, for which this information would be sensitive. This can also become important in certain user-level settings.

\begin{figure}[ht]
 \centering
 \begin{tikzpicture}
 \node[scale=1]{
 \begin{quantikz}
 \qw&\gate[2]{R_{ZZ}(\theta)}&\qw\\
 \qw& &\qw
 \end{quantikz}=
 \begin{quantikz}
 \qw&\ctrl{1}& \qw &\ctrl{1}&\qw\\
 \qw&\targ{}& \gate{R_Z(\theta)} & \targ{}&\qw
 \end{quantikz}
 };
 \end{tikzpicture}
 \caption{\texorpdfstring{Decomposition of the $R_{ZZ}$ gate.}{Decomposition of the R_{ZZ} gate.}}
 \label{fig:rzz-decomposition}
\end{figure}

Not all operations involve sensitive information. For example, in the QAOA discussed, private information appears in the $R_Z$ and $R_{ZZ}$ gates associated with the problem Hamiltonian. By contrast, no private information appears in the $R_X$ gates when their angles are public. Moreover, the $R_{ZZ}$ gates, when decomposed into two $\CNOT$ gates and one $R_Z$ as in Fig. \ref{fig:rzz-decomposition}, show that the entangling $\CNOT$ part does not itself contain the private angle. Therefore, in a QAOA the private content can often be isolated in the single-qubit rotation layer. In general, the gates we are interested in can be efficiently decomposed in this way \cite{Gate_Decomp}. In other cases, such as Grover's algorithm, the information is not contained in rotation angles, but in the circuit structure itself. We therefore distinguish two types of information to be protected in the operations: angle information and structure information. Accordingly, we can also divide the types of operations into public operations, whose description is visible to the server and which can be delegated when they are compatible with the encryption layer, and private operations, which contain confidential information. This public/private distinction is about what the server is allowed to know. Later, when introducing the quantum one-time pad (QOTP) layer, we also use a separate technical distinction between Clifford operations, such as $H$, $S$, $\CNOT$ and $\CZ$, and non-Clifford operations, such as generic rotations, $T$ gates and $\CCZ$.

In particular, on QOTP-encrypted data, public Clifford operations can be tracked by classical Pauli-key updates, while non-Clifford operations require an additional gadget, interaction, decomposition, local client step, or other explicit protocol.

Throughout the paper we use the phase-rotation convention
\begin{equation}
 R_X(\theta)=e^{-i\theta X/2}, \qquad R_Y(\theta)=e^{-i\theta Y/2},
\end{equation}
\begin{equation}
 R_Z(\theta)=e^{-i\theta Z/2}, \qquad R_{ZZ}(\theta)=e^{-i\theta Z\otimes Z/2},
\end{equation}
so that
\begin{equation}
 \CNOT_{12}(I\otimes R_Z(\theta))\CNOT_{12}=R_{ZZ}(\theta).
\end{equation}
Here qubit $1$ is the control and qubit $2$ the target for $\CNOT_{12}$, while $\CZ_{12}$ acts symmetrically on qubits $1$ and $2$.

Distributed quantum-computing settings are relevant when privacy is obtained by splitting information between non-colluding servers or routing nodes, a scenario that also appears in distributed quantum machine learning \cite{Distr_QML,Q_Feder_Learn,Gradient_Hiding}. 

\begin{figure}
 \centering
 \includegraphics[width=\linewidth]{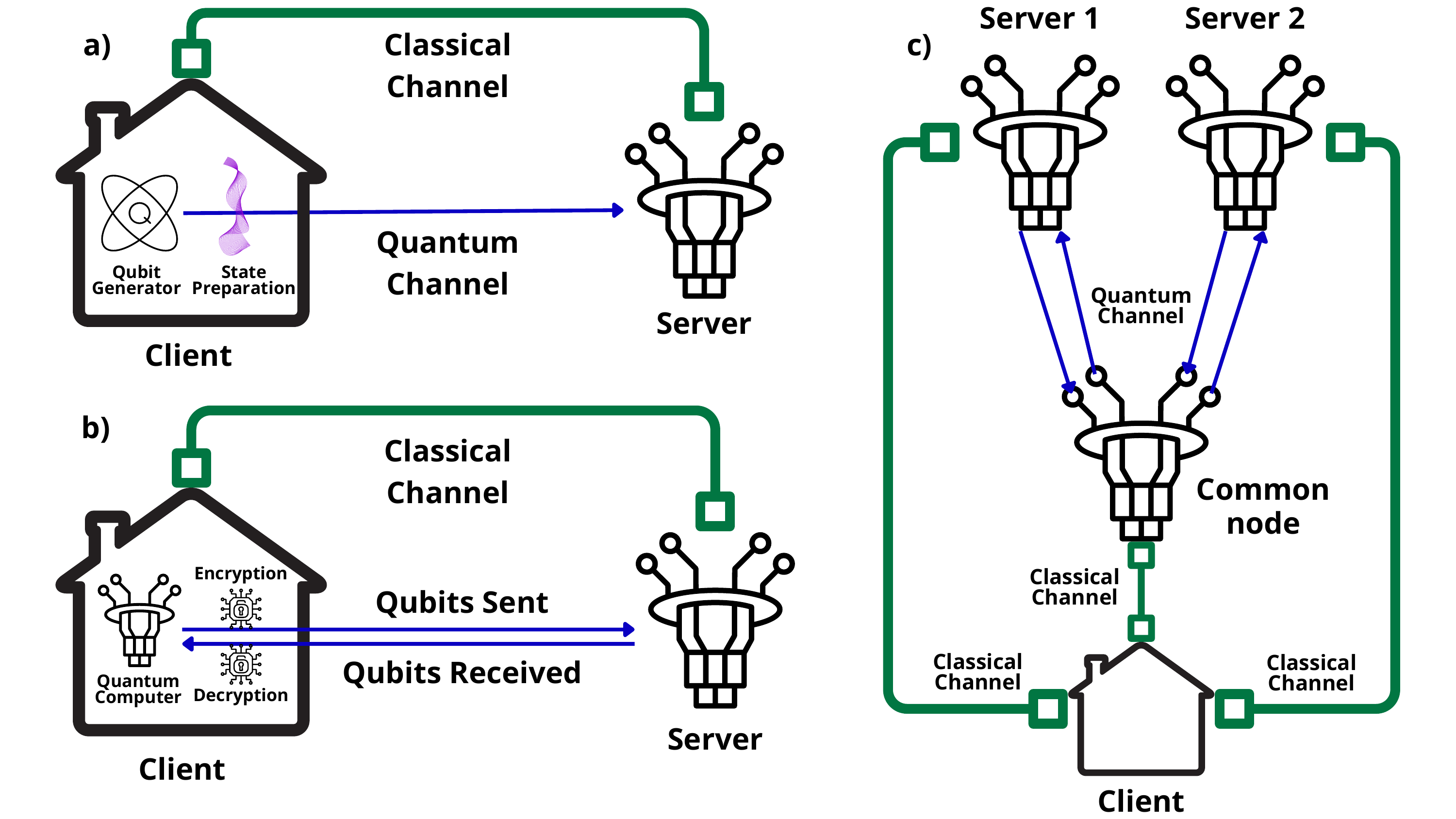}
 \caption{\texorpdfstring{General communication scheme between the client device and the server. a) MBQC: the client sends the qubits to the server and the server performs all quantum operations, which are classically controlled. b) CBQC: the client and the server exchange qubits and perform operations according to the client's instructions. c) Distributed computing: the client coordinates two servers that a priori should not communicate through a common node. In panel (c), the common node is part of the trust/leakage model: it is assumed to be trusted or non-colluding, depending on the security model. Panel (c) represents only the communication and routing architecture. It does not by itself imply blindness or malicious security; the corresponding privacy claim depends on the non-total-collusion, hidden-matching, key-hiding, and side-channel assumptions stated in Protocol \ref{Protocol: 4}.}{General communication scheme between the client device and the server. a) MBQC: the client sends the qubits to the server and the server performs all quantum operations, which are classically controlled. b) CBQC: the client and the server exchange qubits and perform operations according to the client's instructions. c) Distributed computing: the client coordinates two servers that a priori should not communicate through a common node. In panel (c), the common node is part of the trust/leakage model: it is assumed to be trusted or non-colluding, depending on the security model. Panel (c) represents only the communication and routing architecture. It does not by itself imply blindness or malicious security; the corresponding privacy claim depends on the non-total-collusion, hidden-matching, key-hiding, and side-channel assumptions stated in Protocol 4.}}
 \label{fig: General Scheme}
\end{figure}

These ideas lead to two of the main approaches for realizing private delegated quantum computing \cite{Introd}. The first is measurement-based blind quantum computation (MBQC) \cite{MBQC,Universal_Blind}, which uses measurements and corrections on a graph state to perform the computation. The second is circuit-based blind quantum computation (CBQC) \cite{Secure_Assisted,Delegating_private,T_gate}, which encrypts the data with a quantum one-time pad, computes on the encrypted data, and then decrypts it; this is useful as an analogy with BB84-style random Pauli-basis choices \cite{BB84}. Some CBQC variants use restricted encryption layers, including $Z$-based schemes in their specific setting \cite{Z_Encryption}; however, the information-theoretic QOTP privacy used in this paper relies on fresh secret Pauli $X$ and $Z$ masks unless explicitly stated otherwise \cite{T_gate}. Full-blind delegated quantum computation (FBQC) \cite{Full_Blind} extends this second approach by also hiding the operations performed within the protocol transcript, up to the leakage explicitly allowed by that construction. In all these approaches, the client must be able to generate certain quantum states and perform a minimum number of quantum operations, and in some cases must be able to store all the qubits of the process. The communication pattern is shown in Fig. \ref{fig: General Scheme}.

Some of these protocols have been successfully demonstrated experimentally \cite{QC_Encrypted_data,Demonstration}, both for simple cases and for algorithms such as Grover's algorithm \cite{Grover} or the Deutsch-Jozsa algorithm \cite{Deutsch_Jozsa}. They have also been adapted for academic tasks, such as estimating Mahalanobis distance \cite{Mahalanobis}, and industrial applications, such as variational quantum algorithms \cite{Delegated_Variational}, the QAOA \cite{Delegated_QAOA} or quantum neural networks (QNNs) \cite{Delegated_QNN}. Finally, various equivalences between different protocols have been studied \cite{Equivalence}.

This paper uses these techniques as ingredients in a resource-based hierarchy: first a fully capable $M$-qubit client, then independent single-qubit devices, then restricted devices with a finite-grid angle-ambiguity primitive, and finally a separate classical-client branch that changes the trust model.

\paragraph{\texorpdfstring{Contributions and scope.}{Contributions and scope.}}
The contributions of this paper are: (i) a resource-based hierarchy of private delegated quantum computation protocols; (ii) a finite-grid routing-hidden sign-randomized angle-ambiguity technique for certain one-qubit Pauli-axis rotations, $R_X$, $R_Y$ and $R_Z$, under hidden Pauli masks, with transcript leakage and required assumptions stated explicitly; (iii) a classical-client, non-total-collusion protocol based on persistent matching-hidden split-QOTP and shuffled $r$-share sign-randomized angle sharing, with split-mask readout appearing as the final measurement instance of the persistent Pauli frame; and (iv) a trap-based detection layer under stated randomization assumptions. Tables \ref{tab:protocol-summary-resource} and \ref{tab:protocol-summary-security} summarize the intended scope of the protocols.

\noindent The compact protocol summary is split into Tables \ref{tab:protocol-summary-resource} and \ref{tab:protocol-summary-security} to separate resource requirements from security scope. Fresh trusted randomness for QOTP keys is treated as a standing assumption for the QOTP-based branches rather than as a separate protocol branch.

\begin{table*}[t]
\centering
\caption{\texorpdfstring{Resource hierarchy for the protocol branches. Structural privacy depends on the randomized compiler and leakage function.}{Table 1A: Resource hierarchy for the protocol branches. Structural privacy depends on the randomized compiler and leakage function.}}
\scriptsize
\setlength{\tabcolsep}{3pt}
\renewcommand{\arraystretch}{1.12}
\begin{tabular}{@{}L{0.12\textwidth}L{0.23\textwidth}L{0.12\textwidth}L{0.27\textwidth}L{0.20\textwidth}@{}}
\toprule
Item & Client capability & Servers & Delegated operation class & Extra primitive needed \\
\midrule
Protocol \ref{Protocol: 1} & fully capable \(M\)-qubit client with local private gates and measurement & 1 & public Clifford gates on encrypted data; private and non-Clifford gates local if arity \(\leq M\) & decomposition or extra primitives for larger unsupported non-Clifford gates \\
\midrule
Protocol \ref{Protocol: 2} & client with \(M\) independent single-qubit devices plus routing control & 1 & public multi-qubit Clifford gates; local single-qubit gates by the client; non-Clifford support needs a Clifford-plus-local-single-qubit decomposition & randomized routing/padding compiler for structural privacy; restricted-client primitive when local gates are unavailable \\
\midrule
Protocol \ref{Protocol: 3} & restricted single-qubit devices with QOTP keys and routing control & 1 & public Clifford gates plus private Pauli-axis rotations through the finite-grid \(r\)-share sign-randomized primitive & finite-grid \(r\)-share sign-randomized Pauli-axis angle-sharing with hidden-sign, hidden-routing, grouping and side-channel assumptions \\
\midrule
Protocol \ref{Protocol: 4} & classical client coordinating key shares and matchings & two computational servers plus common node & public Clifford operations on persistent matching-hidden split-QOTP data; private Pauli-axis rotations via shuffled \(r\)-share sign-randomized angle sharing; split-frame readout & matching compiler satisfying \(\epsilon_{\mathrm{key}}\)-key hiding; recommended common-node Pauli refresh; hidden aligned final \(X\) mask for readout \\
\bottomrule
\end{tabular}
\label{tab:protocol-summary-resource}
\label{tab:protocol-summary}
\end{table*}

\begin{table*}[t]
\centering
\caption{\texorpdfstring{Security scope of the protocol branches. QOTP state privacy, angle ambiguity or transcript unlinkability, structural privacy, and trap detection are separate guarantees; angle ambiguity is not universal blindness, structural privacy is compiler- and leakage-dependent, and trap detection is not malicious security.}{Table 1B: Security scope of the protocol branches. QOTP state privacy, angle ambiguity or transcript unlinkability, structural privacy, and trap detection are separate guarantees; angle ambiguity is not universal blindness, structural privacy is compiler- and leakage-dependent, and trap detection is not malicious security.}}
\scriptsize
\setlength{\tabcolsep}{3pt}
\renewcommand{\arraystretch}{1.12}
\begin{tabular}{@{}L{0.12\textwidth}L{0.27\textwidth}L{0.19\textwidth}L{0.26\textwidth}L{0.14\textwidth}@{}}
\toprule
Item & Privacy target & Adversarial model & Declared leakage / limitation & Detection status \\
\midrule
Protocol \ref{Protocol: 1} & QOTP state privacy; limited output privacy for Pauli-basis readout & purified semi-honest server for QOTP state privacy; no malicious proof & circuit skeleton, timing and larger non-Clifford gates need decomposition or extra primitives & optional trap-based detection \\
\midrule
Protocol \ref{Protocol: 2} & QOTP state privacy and leakage-dependent structural privacy & purified semi-honest server for QOTP state privacy; side-channel assumptions & routing, timing and public Clifford skeleton may leak structure; local gate set limits non-Clifford support & optional trap-based detection \\
\midrule
Protocol \ref{Protocol: 3} & QOTP state privacy; leakage-dependent structural privacy; transcript-level Pauli-axis angle ambiguity, not universal blindness. & purified semi-honest for QOTP state privacy; transcript-only semi-honest for angle sharing; hidden-sign, hidden-routing, finite-grid masking and side-channel assumptions & visible shares are not logical rotations; grouping, routing, repeated angles, optimizer correlations or side channels can reduce ambiguity; no full blindness theorem & optional trap-based detection \\
\midrule
Protocol \ref{Protocol: 4} & state privacy under persistent split-QOTP plus \(\epsilon_{\mathrm{key}}\)-key hiding; transcript-level angle unlinkability; output privacy under hidden aligned final \(X\) mask. & non-total-collusion purified semi-honest coalitions & common-node matching leakage, side channels, trivial/low-entropy matching, absence of the refresh share if not used, or the full coalition \(\{S_1,S_2,R\}\) defeat or degrade the claim; no malicious security & not malicious secure without verification/authentication \\
\bottomrule
\end{tabular}
\label{tab:protocol-summary-security}
\end{table*}
\FloatBarrier

The structure of this paper is as follows. Sec. \ref{sec:model} fixes the resource, leakage and adversarial model. Sec. \ref{sec: techniques} then reviews the techniques used later in the paper. Sec. \ref{sec: protocols industry} and Sec. \ref{sec: protocols user} adapt them to clients with industry-level and user-level resources, respectively. We then discuss the trap-based detection layer in Sec. \ref{sec: verification}, give concrete algorithmic examples in Sec. \ref{sec: examples}, and close with an explicit summary of security assumptions and limitations in Sec. \ref{sec: limitations} before the conclusions.

\section{\texorpdfstring{Model, resources and privacy notions}{Model, resources and privacy notions}}\label{sec:model}
This section fixes the common model used by the protocol hierarchy. Individual protocols may add stronger requirements or allow additional leakage, but they inherit the distinctions below unless stated otherwise.

\paragraph{\texorpdfstring{Client and server resources.}{Client and server resources.}}
The client may range from a fully capable $M$-qubit device to independent single-qubit devices, restricted single-qubit devices, or a purely classical controller. When the client has quantum hardware, we assume only the capabilities explicitly assigned to that protocol, chosen from preparing qubits, receiving qubits, sending qubits, applying local one-qubit or multi-qubit gates, measuring locally and sampling secret randomness. A fully capable $M$-qubit client can bring at most $M$ qubits into local memory at one time; independent-device clients can act locally on each device but cannot apply local entangling gates; restricted devices may be limited to the gates needed for QOTP masks and routing-dependent primitives.

The server side has enough quantum memory, communication bandwidth and computational capability to execute the delegated operations specified by the protocol. The single-server protocols assume one server with enough memory for the outsourced register. A distributed or multi-server setting is used only when the computation is explicitly split across parties or when a non-collusion assumption is part of the privacy claim. It is therefore not simply an implementation detail of a larger server.

\paragraph{\texorpdfstring{Idealized correctness model.}{Idealized correctness model.}}
For correctness analysis we use ideal quantum channels and ignore physical noise in the stored or transmitted qubits. This ``no quantum errors'' assumption means that noise is not modeled in the correctness proof; it does not mean that an adversarial server is guaranteed to follow the protocol. Communication may take time and may affect the scheduling cost, but in the idealized model it does not introduce physical errors into the quantum states.

\paragraph{\texorpdfstring{Adversarial and leakage model.}{Adversarial and leakage model.}}
Unless a stronger model is explicitly stated, privacy is claimed conditioned on the declared leakage and under the adversarial model assigned to the protocol. In the semi-honest cases, the server applies exactly the prescribed quantum channels, follows the communication schedule, and retains the classical transcript, timing information, public instructions and prescribed quantum systems visible to it. This does not grant the server an unprescribed measurement of every state. In a purified semi-honest model, the server may additionally keep purifying systems, Stinespring environments, auxiliary registers, measurement outcomes, randomness and classical side information compatible with an honest implementation. In a malicious model, the server may apply arbitrary CPTP maps, deviate from the schedule, introduce auxiliary systems, or perform unprescribed measurements. The leakage is formalized by the declared leakage function $L(D)$ below.

\paragraph{\texorpdfstring{Leakage-relative privacy definition.}{Leakage-relative privacy definition.}}
Let $D$ denote the private description of the delegated computation, including the private input description, private gate parameters, hidden structure, routing choices and output masks that are intended to remain hidden. A leakage function $L$ maps $D$ to the declared information that the protocol allows the adversary to learn. Privacy is always relative to $L(D)$, and two computations are compared only when they induce the same declared leakage, that is, when $L(D)=L(D')$.

For a protocol $\Pi$, an adversary $A$, and security parameter or approximation parameter implicit in the implementation, we say that the adversary's view is $\varepsilon$-simulatable relative to $L$ if there exists a simulator $\operatorname{Sim}_{A}$ such that
\[
\Delta\!\left(
\rho_{A,D}^{\Pi},
\operatorname{Sim}_{A}(L(D))
\right)
\leq
\varepsilon,
\qquad
\Delta(\rho,\sigma)=\frac12\|\rho-\sigma\|_1.
\]
Here $\rho_{A,D}^{\Pi}$ is the final classical-quantum view of adversary $A$ in protocol $\Pi$, including the adversary's retained quantum systems and classical side information allowed by the chosen adversarial model. The term $\operatorname{Sim}_{A}(L(D))$ is a simulated view generated using only the declared leakage. Thus a privacy claim says that, conditioned on the declared leakage, the real and simulated adversarial views are indistinguishable up to $\varepsilon$ in trace distance.

\paragraph{\texorpdfstring{Declared leakage.}{Declared leakage.}}
Depending on the protocol, $L(D)$ may include the number of qubits or wires, the public gate alphabet, the number of rounds, message sizes, timing, a public routing template, the precision of the angle grid, a public supergraph or public observable template if one is used, and the number of shots or iterations. Unless explicitly declared, $L(D)$ should not include private angles, QOTP keys, QOTP key shares, output masks, hidden grouping, hidden routing choices, common-node matching permutations, or which visible angles are real, dummy, padding, cancelling, masked-idle or decoy angles.

\paragraph{\texorpdfstring{Transcript-only semi-honest model.}{Transcript-only semi-honest model.}}
In the transcript-only semi-honest model, the party follows the prescribed classical interaction pattern and the claim concerns only the distribution of the visible classical transcript, conditioned on the declared leakage. This model is used only for classical transcript-hiding statements such as finite-grid angle-share ambiguity, sign-randomized angle sharing, hidden grouping, routing ambiguity, transcript-level unlinkability, and dummy, padding, masked-idle and cancelling angle ambiguity. Such statements quantify transcript ambiguity up to matching leakage or structural leakage. They do not by themselves imply full quantum-state privacy or full simulator-based blindness.

\paragraph{\texorpdfstring{Purified semi-honest model.}{Purified semi-honest model.}}
The purified semi-honest model is the default model for QOTP-based state privacy. The server follows the prescribed interface, including the instructed channels, message schedule and public operations, but may keep all purifying systems, Stinespring environments, auxiliary registers, measurement outcomes, randomness and classical side information compatible with an honest implementation. If the prescribed honest operation is a CPTP map
\[
\mathcal E(\rho)
=
\operatorname{Tr}_{E}
\left[
U
\left(
\rho\otimes |0\rangle\langle 0|_{E}
\right)
U^{\dagger}
\right],
\]
then a purified semi-honest server may keep the environment $E$ rather than discarding it. QOTP state-privacy claims for Protocols \ref{Protocol: 1}--\ref{Protocol: 3} should therefore be read as claims about this purified view, conditioned on the declared leakage and on fresh secret Pauli keys.

\paragraph{\texorpdfstring{Malicious adversaries.}{Malicious adversaries.}}
The malicious model is separate. A malicious adversary may apply arbitrary CPTP maps, deviate from the prescribed schedule, introduce auxiliary systems, communicate through unprescribed channels, choose measurements adaptively or return corrupted states. QOTP provides privacy of the encrypted register against an adversary lacking the keys, but it does not provide integrity, correctness or malicious security for the base protocols. Any malicious-security claim requires an additional mechanism and its own proof; it is not inherited from bare QOTP.

\paragraph{\texorpdfstring{Coalitions for Protocol \ref{Protocol: 4}.}{Coalitions for Protocol 4.}}
For the multi-party model used later by Protocol \ref{Protocol: 4}, define the server-side parties as
\[
\mathcal P=\{S_1,S_2,R\},
\]
where $S_1$ and $S_2$ are computational servers and $R$ is the common node. The intended family of allowed coalitions is
\[
\mathcal C_{\mathrm{P4}}=\{C\subsetneq \mathcal P\}.
\]
The full coalition
\[
\{S_1,S_2,R\}
\]
is excluded. In the intended Protocol \ref{Protocol: 4} model, the security resource is split into value shares held by the computational servers and hidden matching information held by the common node. Collusion of $S_1$ and $S_2$ alone is not automatically fatal only when the hidden matching compiler preserves the $\epsilon_{\mathrm{key}}$-key-hiding condition, or when the common-node Pauli refresh supplies an additional hidden uniform share. If the matching is trivial, low-entropy, or leaked through side channels, then the aligned Pauli frame may be reconstructed and the Protocol \ref{Protocol: 4} state-privacy claim degrades accordingly. Claims in this branch should be read under the stated coalition model, conditioned on the declared leakage, and up to matching leakage or structural leakage.

The common node's matching permutations must not be leaked through timing, port identities, physical labels, routing metadata, message sizes or deterministic ordering beyond the declared leakage. If those side channels identify the hidden matching, then the corresponding angle, output or structural privacy claim must be weakened or the leaked information must be added to $L(D)$.

\paragraph{\texorpdfstring{Privacy terminology.}{Privacy terminology.}}
We use \emph{state privacy} for hiding the encrypted quantum state from a server that lacks the QOTP keys, \emph{leakage-dependent transcript-level angle ambiguity} for finite-grid rotation-share transcripts that leave multiple private angle labels possible under the declared leakage, \emph{structural privacy} for hiding the circuit layout, gate positions, routing or computation graph, and \emph{output privacy} for hiding measurement outcomes before the client reconstructs them. These are separate guarantees: a protocol can hide the quantum state while leaking the public circuit skeleton, or hide one angle share while leaking the way shares are grouped. The privacy guarantees are therefore protocol-dependent, leakage-dependent and interpreted through the leakage-relative definition above. A claim should always be read together with the stated client capabilities, delegated gate set, declared leakage function and adversarial assumptions.
For Protocols \ref{Protocol: 1}--\ref{Protocol: 3}, the QOTP state-privacy statements are statements about the server's purified semi-honest quantum view under fresh secret Pauli keys, conditioned on the declared leakage. Protocol \ref{Protocol: 4} is different: its state-privacy claim is state privacy under persistent split-QOTP plus \(\epsilon_{\mathrm{key}}\) key hiding, rather than client-side QOTP, while its classical value shares, hidden matching information and final readout masks are protected only under the stated coalition model and side-channel assumptions. Its angle claim is transcript-level angle unlinkability, leakage-relative, not universal blindness, and not malicious security.

\begin{table}[ht]
\centering
\caption{Adversarial models and guarantees considered in this work.}
\small
\setlength{\tabcolsep}{3pt}
\renewcommand{\arraystretch}{1.12}
\begin{tabular}{p{0.24\linewidth}p{0.30\linewidth}p{0.37\linewidth}}
\toprule
Protocol / primitive & Default adversarial model & Main guarantee \\
\midrule
QOTP state encryption
& Purified semi-honest
& State privacy against views missing the aligned Pauli key \\
Protocol \ref{Protocol: 1}
& Purified semi-honest server
& QOTP state privacy; public Clifford-frame delegation \\
Protocol \ref{Protocol: 2}
& Purified semi-honest server
& QOTP state privacy; compiler-relative structural privacy \\
Protocol \ref{Protocol: 3}
& Purified semi-honest for state; transcript-only for angle sharing
& QOTP state privacy plus transcript-level Pauli-axis angle ambiguity, not universal blindness \\
Sign-randomized angle sharing
& Transcript-only semi-honest
& Finite-grid transcript ambiguity under hidden signs, grouping and routing \\
Protocol \ref{Protocol: 4}
& Non-total-collusion purified semi-honest coalitions
& State privacy under persistent split-QOTP plus \(\epsilon_{\mathrm{key}}\) key hiding; transcript-level angle unlinkability \\
Common-node shuffling
& Side-channel-controlled common node under non-total-collusion
& Hidden matching of QOTP shares, angle shares and auxiliary slots \\
Randomized compiler
& Leakage-relative semi-honest
& Structural privacy conditioned on the declared leakage \\
Dummy / padding / decoy angles
& Transcript-only semi-honest
& Ambiguity amplification; no verifier semantics at this stage \\
Embedded verifier layer
& Malicious deviations under hidden-location assumptions
& Statistical detection of some deviations, introduced later \\
Authenticated QOTP layer
& Malicious adversaries
& Integrity / tamper detection, if separately instantiated \\
\bottomrule
\end{tabular}
\label{tab:adversarial-models}
\end{table}

\section{\texorpdfstring{Delegated quantum computation techniques}{Delegated quantum computation techniques}}\label{sec: techniques}
This section collects the tools used by the protocol hierarchy. MBQC and FBQC are included mainly as background and comparison points, while QOTP privacy, Clifford key updates, structural-view indistinguishability and the routing-hidden Pauli-axis rotation primitive are the tools used directly later.
\subsection{MBQC}
Measurement-based quantum computing is based on preparing an entangled quantum state, called a graph state, and on performing a sequence of conditional qubit measurements. This method can realize the same computations as the quantum circuit formalism, with entanglement used only in the state construction. Moreover, since the results of each measurement step are random, one applies corrections to later measurement choices based on earlier outcomes. We use MBQC only as a reference point for blind and trap-based ideas, not as the main implementation path of the protocol hierarchy.

The graph state is represented by a graph, where each node is a qubit and each edge is an entangling operation between the qubits it links. The graph state contains ancilla qubits and output-state qubits. The ancilla qubits are measured to form the desired state in the output register, since the entanglement translates modifications in the ancilla qubits into modifications in the output state. All ancilla qubits are prepared in a uniform superposition, and the entanglement is generated with $\CZ$ gates.

\begin{figure}
 \centering
 \begin{tikzpicture}
 [scale=.5,auto=left,every node/.style={draw, circle}]
 \node (n1) at (0,0) {+};
 \node (n2) at (2,0) {+};
 \node (n3) at (4,0) {+};
 \node (n4) at (6,0) {+};
 \node (n5) at (8,0) {+};
 \node (n6) at (0,2) {+};
 \node (n7) at (2,2) {+};
 \node (n8) at (4,2) {+};
 \node (n9) at (6,2) {+};
 \node (n10) at (8,2) {+};

 \foreach \from/\to in {n1/n2,n2/n3,n3/n4,n4/n5,n6/n7,n7/n8,n8/n9,n9/n10,n1/n6,n2/n7,n3/n8,n4/n9,n5/n10}
 \draw (\from) -- (\to);
\end{tikzpicture}
\begin{tikzpicture}
 [scale=.5,auto=left,every node/.style={draw, circle}]
 \node (n1) at (0,0) {$\alpha_1$};
 \node (n2) at (2,0) {$\alpha_2$};
 \node (n3) at (4,0) {$\alpha_3$};
 \node (n4) at (6,0) {$\alpha_4$};
 \node (n5) at (8,0) {$\alpha_5$};
 \node (n6) at (0,2) {$\alpha_6$};
 \node (n7) at (2,2) {$\alpha_7$};
 \node (n8) at (4,2) {$\alpha_8$};
 \node (n9) at (6,2) {$\alpha_9$};
 \node (n10) at (8,2) {$\alpha_{10}$};

 \foreach \from/\to in {n1/n2,n2/n3,n3/n4,n4/n5,n6/n7,n7/n8,n8/n9,n9/n10,n1/n6,n2/n7,n3/n8,n4/n9,n5/n10}
 \draw (\from) -- (\to);
\end{tikzpicture}
 \caption{\texorpdfstring{Generic graph state and graph state used in the method of \cite{Universal_Blind}.}{Generic graph state and graph state used in the method of Universal Blind.}}
 \label{fig: Graph state}
\end{figure}

The key to the method is that measuring a uniform superposition state in one direction of the $X-Y$ plane is equivalent to projecting it in that direction. Operationally, this acts like applying a gate to the state. The associated measurement gate \cite{Measurement} is
\begin{equation}
 J(\theta)=H R_Z(\theta)=\frac{1}{\sqrt{2}}\begin{pmatrix}
 e^{-i\theta/2}&e^{i\theta/2}\\
 e^{-i\theta/2}&-e^{i\theta/2}
 \end{pmatrix},
\end{equation}
which is achieved by measuring at the $\theta$ angle and correcting the state. Up to the physically irrelevant global phase $e^{-i\theta/2}$, this is the familiar convention
\[
 J(\theta)\sim \frac{1}{\sqrt{2}}\begin{pmatrix}
 1&e^{i\theta}\\
 1&-e^{i\theta}
 \end{pmatrix}.
\]
Furthermore, any quantum circuit can be constructed from $\CZ$ and $J(\theta)$ gates.

Once the graph state is generated, no more complex operations need to be applied to it. Moreover, the only multi-qubit operations applied are those used at the beginning of the process. This does not by itself reveal circuit information when the graph pattern is fixed, public or universal, as in universal graph-state constructions. If instead the graph topology is adapted to the target circuit, then the topology itself may leak circuit structure and should be included in the declared leakage or hidden by a universal or padded graph pattern.

Given these characteristics, \cite{Universal_Blind} studies the possibility of using this technique for private delegated computation. In blind MBQC, the client sends randomly encoded qubits and masked measurement instructions; the adapted and randomized measurement angles hide the underlying logical angles up to the leakage allowed by the blind-MBQC construction. This protocol can also be adapted to distributed quantum computing \cite{Distributed_MBQC}, in which several servers store and measure the global graph state.

This protocol can include a trap-qubit check for whether the server is being honest and following the requested computation \cite{Verificar}. For this purpose, a set of trap qubits is inserted, giving deterministic results known only to the client. If an unexpected result is obtained, this indicates that the server is not performing the designated operations.

Its main limitation for the present hierarchy is resource overhead: the number of graph-state qubits depends on the computation size and depth. A client with only $M$ qubits could in principle send graph-state qubits in packets, but this remains an MBQC resource-state approach rather than the QOTP-based circuit hierarchy developed below.

\subsection{CBQC}\label{sec: CBQC}
Circuit-based blind quantum computation techniques execute a quantum circuit while the quantum state is encrypted with a one-time key. At the level of motivation, this is analogous to classical homomorphic encryption \cite{Homomorphic}: the server is asked to operate on encrypted data, while only the client holds the secret information needed to interpret the result. Conceptually, it can be viewed as a realization of the BB84 protocol, where the sender and the receiver of the state are both the client, while the eavesdropper is the server. In this context, the client can only perform $X$, $Z$ and, when two local qubits are available, actual two-qubit $\SWAP$ gates, in addition to generating arbitrary single-qubit states and being able to store them, so it cannot perform universal quantum computation. Each time the client wants to perform a quantum gate that it cannot apply locally, it encrypts the target qubits and sends them to the server for implementation. This encryption is performed by applying an $X^xZ^z$ operation to the target qubit, with the pair $(x,z)\in\{0,1\}^2$ being the random binary key stored by the client. After receiving the qubits back, the client updates the Pauli frame and decrypts according to the relevant rules presented in \cite{Secure_Assisted,Delegating_private}.

The main limitation of this scheme is that the client must be able to store all the qubits of the circuit. In addition, the protocol requires many qubit transmissions and may reveal the operations to the server, which can be problematic in cases such as QAOA.

\subsection{QOTP privacy and delegated Clifford operations}
The circuit-based protocols below repeatedly use the QOTP. For $n$ qubits, and bit strings $x,z\in\{0,1\}^n$, we write
\begin{equation}
 P_{x,z}=\bigotimes_{j=1}^n X^{x_j}Z^{z_j}.
\end{equation}
The client encrypts $\rho$ as $P_{x,z}\rho P_{x,z}^\dagger$ and keeps $(x,z)$ secret.

\begin{lemma}[Information-theoretic privacy of the QOTP]\label{lem:qotp}
For every single-qubit state $\rho$,
\begin{equation}
 \frac{1}{4}\sum_{x,z\in\{0,1\}}X^xZ^z\rho Z^zX^x=\frac{I}{2}.
\end{equation}
More generally, for every $n$-qubit state $\rho$,
\begin{equation}
 \frac{1}{4^n}\sum_{x,z\in\{0,1\}^n}P_{x,z}\rho P_{x,z}^\dagger=\frac{I}{2^n}.
\end{equation}
Therefore, if the Pauli keys are uniformly random, secret from the server, and freshly sampled for each encrypted register, then the server's reduced density matrix is maximally mixed and reveals no information about the encrypted quantum state itself \cite{Secure_Assisted,Private_Quantum_Channels}. Correlated, leaked or reused keys can invalidate this conclusion.
\end{lemma}

\begin{proposition}[Encrypted evaluation of public Clifford gates]\label{prop:clifford-update}
Let $C$ be a Clifford operation on $n$ qubits. Then there exists an efficiently computable key-update function $f_C$ and a phase $\omega(C,x,z)$ such that
\begin{equation}
 CP_{x,z}=\omega(C,x,z)P_{f_C(x,z)}C.
\end{equation}
For the single-qubit key convention $X^xZ^z$, the update rules are
\begin{align}
 H: \ &(x,z)\mapsto (z,x),\\
 S: \ &(x,z)\mapsto (x,z\oplus x),
\end{align}
For $\CNOT_{c,t}$, the update rules are
\begin{align}
 x_c'&=x_c, & z_c'&=z_c\oplus z_t,\\
 x_t'&=x_c\oplus x_t, & z_t'&=z_t.
\end{align}
For $\CZ_{12}$, they are
\begin{align}
 x'_1&=x_1, & x'_2&=x_2,\\
 z'_1&=z_1\oplus x_2, & z'_2&=z_2\oplus x_1.
\end{align}
Equivalently, under $\CZ_{12}$ the $Z$ keys update as $z'_1 = z_1 \oplus x_2$ and $z'_2 = z_2 \oplus x_1$.
Hence public Clifford operations can be delegated on QOTP-encrypted data, with the client updating only a classical Pauli frame.
\end{proposition}

\begin{assumption}[Restriction for encrypted public evaluation]\label{ass:clifford-only}
Whenever the server is asked to apply public operations directly to QOTP-encrypted data, we restrict those operations to the Clifford subgroup unless an additional mechanism for non-Clifford encrypted evaluation is explicitly provided. In particular, $\CCZ$ is non-Clifford and satisfies
\begin{equation}
 \CCZ X_1 = X_1 \CZ_{23}\CCZ,
\end{equation}
so a simple Pauli-key update is no longer sufficient.
Operationally, Clifford gates update the Pauli frame classically, whereas $T$ gates or arbitrary non-Clifford operations generally require local execution by the client, magic-state or gadget interaction, QHE with non-Clifford overhead, encrypted auxiliary magic states, or a blind-computation construction, because they do not normalize the Pauli group.
\end{assumption}

\paragraph{Server view and structural privacy.}
For the structure-hiding discussion, we define the server's view $\mathsf{View}$ to include the number of qubits, the sequence of requested public operations, the physical ports used, timing information, public gates applied by the server, measurement requests and outcomes visible to the server, and every classical message sent by the client. Let $\mathcal{L}$ denote the declared leakage extracted from this view. Structural privacy is obtained only relative to this declared leakage and to the randomized compiler used to generate the delegated instance. The point of the following criterion is therefore to state a sufficient indistinguishability condition, not to prove that every padding or routing strategy automatically satisfies it.

\begin{criterion}[Sufficient structural-privacy criterion]\label{prop:structural-privacy}
Let $\mathcal{C}$ and $\mathcal{C}'$ be two private circuit components with the same declared leakage $\mathcal{L}$. Suppose the client's padding, decoy-gate placement, routing permutations and port randomization, as generated by a specified randomized compiler, induce identical distributions for $\mathsf{View}$ when the hidden component is $\mathcal{C}$ or $\mathcal{C}'$, conditioned on $\mathcal{L}$. Then the server's view distinguishes $\mathcal{C}$ from $\mathcal{C}'$ only through the declared leakage $\mathcal{L}$.
\end{criterion}

The criterion separates three ingredients: the declared leakage $\mathcal{L}$, the concrete randomized compilation procedure, and the resulting indistinguishability of server views. In the protocols below, any structural-privacy statement should be read as conditional on these ingredients being supplied and verified for the relevant circuit family.

When the location pattern of multi-qubit Clifford or Pauli-axis operations is private, the compiler may optionally harden the schedule by padding each visible operation class with dummy or cancelling operations of the same type on dummy or masked-idle registers. These dummy or cancelling operations must go through the same routing, timing, key-update and side-channel model as real operations; otherwise their placement may itself become distinguishable. For Pauli-axis rotations, dummy and cancelling targets should be treated as masked-idle or cancelling targets and compiled through the restricted Pauli-axis primitive of Sec. \ref{sec:restricted-rz}. For multi-qubit Clifford operations, dummy and cancelling operations should use the same visible gate class and should not be trivially identifiable by timing, adjacency, qubit labels, missing key updates or simplified classical metadata. After such a layer, the common node or routing layer should rerandomize the matching so that computational servers see only a padded schedule of operations of the same public type. This is a structural-leakage hardening technique, not a full-blindness claim.

\paragraph{\texorpdfstring{Concrete example: a padded QAOA/QUBO compiler.}{Concrete example: a padded QAOA/QUBO compiler.}}
This compiler is an illustrative sufficient construction, not the only way to obtain structural privacy. Consider a private QUBO/QAOA instance whose cost graph is $G=(V,E)$. The client chooses or publishes a supergraph $G_{\mathrm{pub}}=(V,E_{\mathrm{pub}})$ with $E\subseteq E_{\mathrm{pub}}$. The declared structural leakage contains $G_{\mathrm{pub}}$, $N=|V|$, the QAOA depth and layer schedule, the public gate calendar, the angular precision parameter, and the number of rounds or shots. It does not contain the private edge set $E$.
\begin{equation}
 \CNOT_{u,v}\; R_Z(\theta_{uv})\; \CNOT_{u,v},
 \qquad (u,v)\in E_{\mathrm{pub}}.
\end{equation}
For every public candidate edge, the compiler emits the same public entangling pattern and uses
\begin{equation}
 \theta_{uv}
 =
 \begin{cases}
 \theta^{\mathrm{real}}_{uv}, & (u,v)\in E,\\
 0, & (u,v)\notin E.
 \end{cases}
\end{equation}
The zero label is not sent as a recognizable bare zero rotation. Instead, using the same finite angular grid and precision model as the angle-ambiguity primitive, with $q=2^p$, the compiler represents $\theta_{uv}$ and the fresh mask $\eta_{uv}$ as labels in $\mathbb{Z}_q$ with additions modulo $q$, equivalently as physical angles $2\pi \eta_{uv}/q$, and first writes each single-qubit phase rotation as the logical masked pair
\begin{equation}
 R_Z(-2\pi\eta_{uv}/q)R_Z(2\pi(\theta_{uv}+\eta_{uv})/q),
 \qquad \eta_{uv}\leftarrow\mathbb{Z}_q.
\end{equation}
Equivalently, the physical mask angle is the fractional angle $2\pi\eta_{uv}/q$. All labels are computed modulo $q$, and a fresh secret label $\eta_{uv}$ is used for each compiled occurrence. This masked pair is a logical decomposition, not a prescription to emit two server-visible rotations that remain identifiable as belonging to the same candidate edge. If the server could associate both target rotations and recover their effective signed sums, their net angle would be $\theta_{uv}$, so a dummy edge with $\theta_{uv}=0$ could become distinguishable.

In the intended compiler, the two target labels $-\eta_{uv}$ and $\theta_{uv}+\eta_{uv}$ are treated as independent private Pauli-axis rotation targets. Each target is separately decomposed into $r$ finite-grid shares using the routing-hidden sign-randomized primitive of Sec. \ref{sec:restricted-rz}. The resulting visible shares are embedded in the layer-level routing and shuffling procedure, together with real, dummy, padding, masked-idle and cancelling slots of the same visible operation class. Thus the server does not receive the two masked target rotations as a recognizable pair; it receives only visible grid shares, public timing and physical routing information, while the effective signs and logical groupings are hidden by the QOTP Pauli frame and the routing layer.

Thus real and dummy candidate edges induce the same public topology, the same visible operation classes, the same timing, and the same interaction pattern, while the private edge set $E$ is hidden among the public candidate edges $E_{\mathrm{pub}}$, provided that real and dummy edges induce indistinguishable server-view distributions. The mask $\eta_{uv}$ alone does not provide structural or leakage-dependent transcript-level angle ambiguity; the privacy statement is conditional on the full hidden-sign and hidden-grouping primitive and on the side-channel assumptions stated there.
This compiler realizes structural privacy only relative to the declared leakage: it does not hide $G_{\mathrm{pub}}$, the problem size, the QAOA depth, the number of layers, the precision parameter, or the number of shots. In repeated shots or adaptive optimization, fresh randomness per shot is still not enough if timing, physical port labels, parameter sequences, stopping rules, deterministic batch identities, loss patterns, visible key-update metadata, optimizer trajectories, or correlations between iterations are visible and depend on the private instance. Those quantities must either be included in the declared leakage or padded and randomized until the induced server-view distributions remain indistinguishable.

\paragraph{\texorpdfstring{Reusable angle-ambiguity primitive.}{Reusable angle-ambiguity primitive.}}
For restricted clients that cannot apply private one-qubit Pauli-axis rotations locally, Sec. \ref{sec: protocols user} uses a finite-grid routing-hidden sign-randomized Pauli-axis rotation sharing primitive. Conceptually, the server sees visible grid rotation shares, while hidden QOTP Pauli keys choose the logical signs and hidden routing prevents the server from reliably grouping the shares of one logical angle. The target angle is first compiled to a public finite angular grid; the private value is the resulting grid label, not the existence of the grid or its precision parameter. This is a leakage-dependent angle-ambiguity primitive under hidden-sign, hidden-grouping, finite-grid masking and side-channel assumptions; it is not a general blind-computation proof or a generic non-Clifford homomorphic-evaluation method.

\subsection{FBQC}
Full-blind delegated quantum computation is another comparison point: it hides the quantum data through a one-time pad and obscures the delegated gate choice within a prescribed computation pattern \cite{Full_Blind}. The operations are hidden through the use of nine qubits, so that each subset of them is associated with a specific operation. Instead of revealing a single selected operation, the protocol embeds the possible basic operations in the same prescribed pattern. However, only the client knows which qubits are of interest, and therefore, which operation is effectively selected. For the present paper, the important point is that this stronger hiding comes with substantial qubit and transmission overhead, so it is treated as background rather than as the default construction.

\paragraph{Base convention for the QOTP-based protocols.}
For the QOTP-based protocols introduced below, the server is only asked to apply public Clifford operations directly to QOTP-encrypted data. This convention avoids the key-dependent non-Pauli byproducts that arise when a non-Clifford gate is applied to an encrypted register. Any private operation, and in particular any non-Clifford operation, is instead performed locally by the client whenever the client has the necessary local capability. If such local execution is impossible, then the operation must be decomposed into admissible layers or implemented by an additional primitive, which is outside the base protocol.

For Protocols \ref{Protocol: 2} and \ref{Protocol: 3}, a more expressive but resource-heavier variant may incorporate known gadget-based or QHE-style techniques for non-Clifford gates, provided that the client can supply the required auxiliary states or interaction. We do not use this route in Protocol \ref{Protocol: 4}, where the client-side capabilities are intentionally below that threshold.

For a client made of independent universal single-qubit devices, the admissible target circuits are those that can be compiled as
\begin{equation}
 U_{\mathrm{target}}
 =
 C_LD_LC_{L-1}D_{L-1}\cdots C_1D_1C_0,
\end{equation}
where each $C_j$ is a public Clifford circuit delegated to the server and each
\begin{equation}
 D_j=\bigotimes_i d_{j,i}
\end{equation}
is a product of single-qubit gates performed locally by the client. For a fully capable $M$-qubit client, a non-Clifford gate $U$ of arity $r\leq M$ can instead be handled by the local sequence
\[
 \text{receive affected qubits}
 \longrightarrow
 \text{decrypt}
 \longrightarrow
 U
 \longrightarrow
 \text{freshly encrypt}
 \longrightarrow
 \text{return to server}.
\]
Equivalently, on the affected register,
\begin{equation}
 P_{x,z}\rho P_{x,z}^{\dagger}
 \longrightarrow
 \rho
 \longrightarrow
 U\rho U^{\dagger}
 \longrightarrow
 P_{x',z'}U\rho U^{\dagger}P_{x',z'}^{\dagger}.
\end{equation}
The cost of this conservative compilation strategy is an increase in gate count, communication rounds, scheduling complexity and re-encryption steps. A multi-qubit non-Clifford gate that is decomposed into public Clifford layers and local single-qubit gates may require several client-server handoffs. This overhead is a resource tradeoff used to avoid assuming encrypted non-Clifford evaluation by the server.

\begin{remark}[\texorpdfstring{Direct non-Clifford delegation is not a base mechanism}{Direct non-Clifford delegation is not a base mechanism}]
A directly delegated $T_Z=R_Z(\pi/4)$ on QOTP-encrypted data is a special case in which the key-dependent byproduct is a single-qubit Clifford correction. A client capable of applying $S$ can in principle correct this byproduct locally after the qubit is returned. However, the base protocols do not rely on this direct non-Clifford delegation path. Instead, $T$ and $T^\dagger$ gates are treated as local single-qubit gates whenever the client can implement them. If the client is restricted to $X,Z$, even this local or corrective strategy is unavailable.

Similarly, direct $\CCZ$ evaluation is not part of the base protocol. Applying $\CCZ$ directly to QOTP-encrypted inputs produces key-dependent entangling Clifford byproducts. For hidden $X$-keys $a_1,a_2,a_3$, the byproduct has the form
\begin{equation}
 \Gamma(a_1,a_2,a_3)
 =
 \CZ_{23}^{a_1}
 \CZ_{13}^{a_2}
 \CZ_{12}^{a_3}
 Z_3^{a_1a_2}
 Z_2^{a_1a_3}
 Z_1^{a_2a_3}.
\end{equation}
A fully capable $M$-qubit client could correct such byproducts only if the affected qubits can be brought into local memory and the client can apply the required $\CZ$ corrections. This direct-delegation route is therefore not available to a client consisting only of independent single-qubit devices. In the base solution, Protocol \ref{Protocol: 1} performs a three-qubit $\CCZ$ locally when $M\geq 3$, while Protocol \ref{Protocol: 2} must use a Clifford-plus-local-single-qubit decomposition when the client has universal single-qubit devices.
\end{remark}

\section{\texorpdfstring{Protocols for industry-scale clients}{Protocols for industry-scale clients}}\label{sec: protocols industry}
This section analyzes cases in which the client is an organization with larger quantum resources and problem sizes. First, in Sec. \ref{sec: M total} we study the unconstrained case, where the main limitation is memory size; this serves as an initial scheme for the later protocols. Then, in Sec. \ref{sec: M qubit} we address the case in which the client's allowed operations are constrained. Specifically, the client cannot perform multi-qubit operations: instead of having one $M$-qubit quantum computer, the client has multiple independent single-qubit devices.

\subsection{\texorpdfstring{Fully capable $M$-qubit client}{Fully capable M-qubit client}}\label{sec: M total}
The first main delegated-computation protocol in the hierarchy is the fully capable $M$-qubit client. As a simple initial case, consider a company that wants to perform a quantum computation, has an $M$-qubit quantum computer that can perform all available quantum operations (including multi-qubit operations and measurements), but needs to perform a computation with $N$ qubits, such that $N>M$. In this case, the main limitation faced by this client is that it cannot store all qubits simultaneously, so it needs server-side quantum memory. The server is used as delegated quantum memory and as an evaluator of public Clifford layers, while the client keeps private and non-Clifford operations local whenever the affected register fits in its $M$ qubits.

Operationally, the server acts as quantum memory: the client sends it the qubits on which it is not operating at that moment, while keeping in the client's device the qubits on which it has to perform operations.

The data embedded in the states sent to the server must be information-theoretically hidden, so the client applies one-time pad encryption as explained in Sec. \ref{sec: CBQC}. When the client sends a qubit to the server, it applies an $X^xZ^z$ operation to that qubit with the secret key $(x,z)$ and then undoes it when the qubit is returned, using the methods of \cite{Homomorphic_Circuits,Z_Encryption,Full_Blind,Secure_Assisted}. For the generation of the random keys for encryption, one option is to use idle qubit slots to generate fresh random key bits; two independent bits are needed for each single-qubit QOTP key $(x,z)$. Another option is to generate all the necessary keys prior to the execution of the algorithm and thus be able to perform the protocol with all the qubits. However, this is not possible in all algorithms, such as in an HHL algorithm \cite{HHL}, since we do not know in advance how many repetitions of the circuit must be performed, and therefore, the length of the final circuit. The third option is to use an external private randomness source, provided it is trusted, sufficiently unbiased for the required security level, secret from the server, and never reused in a way that correlates QOTP keys. The local randomness primitive below can be run whenever the client has free qubit slots after sending a round of encrypted qubits to the server; it is not itself a delegated-computation protocol.
\begin{primitive}[Randomness primitive for QOTP keys]\label{Primitive: 1}
$ $

\textbf{Output:} Random classical bits for Pauli-key generation.
 \begin{enumerate}
 \item The client prepares a trusted $\ket{+}$ state in each free slot it has and measures it in the computational basis to obtain one fresh random classical bit per measured qubit.
 \item Repeat step 1 until the client has enough independent bits for all required QOTP keys.
 \end{enumerate}
\end{primitive}

\paragraph{Primitive 1: assumptions and limitations.}
$ $

\textbf{Parties.} Client only.

\textbf{Assumptions.} The state preparation and measurements used for key generation are trusted, and the generated bits are used as fresh secret Pauli-key bits. This primitive assumes trusted $\ket{+}$ preparation and computational-basis measurement, or an equivalent private randomness source. In particular, a client restricted to Pauli $X,Z$ operations on $\ket{0}$ cannot generate $\ket{+}$ states by itself.

\textbf{Correctness.} Measuring $\ket{+}$ in the computational basis yields an unbiased bit, so repeating the procedure produces uniformly random classical key bits.

\textbf{Limitation.} If the randomness source is biased, leaked or reused across different encrypted registers, the privacy guarantee of Lemma \ref{lem:qotp} no longer follows.

Because this client has a richer quantum device, one could consider other encryption layers, including multi-qubit gates when several qubits are sent at once. Here we restrict the base construction to the $X$ and $Z$ one-time pad because it gives the privacy statement of Lemma \ref{lem:qotp} and because the decryption rules for public Clifford operations are explicit. Since the server is also a quantum computer that can perform operations, it is natural to use it in parallel with the client. However, operations on encrypted states make the decryption rules more complicated. In particular, by Assumption \ref{ass:clifford-only}, we only allow the server to apply public Clifford operations directly on QOTP-encrypted states unless an additional non-Clifford mechanism is supplied. Thus, in the base protocol, the server performs only compatible public Clifford operations on encrypted data, while the client performs all private operations and all non-Clifford operations whose affected register fits in its local memory.

Therefore, Protocol \ref{Protocol: 1} proceeds through several steps. In the first step, the client iteratively generates $Q$ groups of up to $M$ qubits in $Q$ rounds, where $Q=\lceil N/M \rceil$, each generated qubit representing one logical qubit of the target circuit. In each round, the client applies all currently available circuit operations until one of those qubits must interact with a qubit from another group, at which point the client sends the group of qubits to the server. Before sending these qubits, the client encrypts them with the one-time pad, applying the encryption $X^{x_i}Z^{z_i}$ to qubit $i$, so that its key is $(x_i,z_i)$. The protocol uses commuting operations as much as possible and stops only when necessary. The client repeats this until it has generated $(Q-1)M$ qubits. In each round, if the server can perform public Clifford operations on the encrypted qubits it is storing, it can do so, reducing the client's future operations load through the Pauli-frame update rules.

When the client reaches the last group, of $N-(Q-1)M$ qubits, the client keeps from the previous group the $P$ qubits that will interact with this last group, with a maximum $P$ such that the total number of qubits in the client is $M$. After that, the client generates the qubits of this group and asks the server for $M-P-N+(Q-1)M=QM-N-P$ stored qubits that must interact with the qubits currently held by the client. This optimizes the number of future requests. The client then applies all the corresponding operations on these qubits.

The second step is for the client to determine which qubits are needed next to continue the circuit process and which are no longer needed from among those currently stored. Then, the client encrypts the qubits it does not need with the one-time pad and sends them to the server, requesting in return the qubits it does need and then decrypting them. After decrypting them, the client operates on them as far as possible and repeats this second step until the whole circuit is finished. During this whole process, the server performs the compatible public Clifford operations it can with the encrypted qubits it has; a private or non-Clifford operation $U$ acting on $r\leq M$ qubits is executed locally by requesting the affected encrypted qubits, decrypting, applying $U$, freshly encrypting, and returning them. Operations of arity larger than $M$ must be decomposed into smaller admissible operations or implemented by an additional primitive.

It is important to emphasize that the client determines which operations can be delegated and when, since the client communicates the public operations only when they are needed. In addition, these operations, when applied to encrypted qubits, require the corresponding Pauli-frame update or decryption rule afterwards, so this can limit the number and types of operations that the server can perform.

The final step is to perform the necessary measurements of the circuit, for which the client asks for the groups of qubits needed and measures them locally. The client can also ask the server for help with Pauli-basis measurements. In the computational basis, the hidden $X$ key flips the reported bit and the hidden $Z$ key does not affect it, so the client can correct the outcome classically as long as the corresponding keys remain secret. For arbitrary measurement bases, this is not automatic: the effective basis depends on the hidden Pauli, so the client must either adapt the requested basis or use an explicit measurement subprotocol. The client and the server can start measuring the qubits with no remaining operations after their last associated gate, so that measurement steps are mixed with operation steps, saving extra requests.

Therefore, the proposal is the following protocol.

\begin{protocol}\label{Protocol: 1}
$ $
 
 \textbf{Input}: Quantum circuit, private/public operations list, $N$ (qubits needed) and $M$ (qubits of the client).

 \textbf{Output}: Result of the circuit execution.
 \begin{enumerate}
 \item Let $Q=\lceil N/M\rceil$ be the number of client batches required to cover the $N$ logical qubits with an $M$-qubit client register.
 \item For $i=0,\ldots,Q-2$:
 \begin{enumerate}
 \item Generate a group of $M$ qubits, from qubit $iM$ to qubit $(i+1)M-1$.
 \item Apply circuit gates until other qubits are needed.
 \item Encrypt the qubits by applying $X^{x_j}Z^{z_j}$ to qubit $j$, with fresh random keys $x_j,z_j\in \{0,1\}$.
 \item Send the group of qubits to the server.
 \item (Parallel) Check if the server can execute any compatible public Clifford operation with the encrypted qubits it has. If so, the server is instructed to perform it and the client updates the Pauli frame.
 \end{enumerate}
 \item Last round:
 \begin{enumerate}
 \item In the previous round, keep any qubits that may interact with the qubits still to be generated.
 \item Generate the last qubits, from qubit $(Q-1)M$ to qubit $N-1$.
 \item If the client currently holds $M'<M$ qubits, it asks the server for $M-M'$ qubits that must interact with those held by the client. The client decrypts the qubits it receives with the key $(x_j,z_j)$ of the corresponding qubit $j$ and the appropriate Pauli-frame update rules for the operations applied while the qubit was stored by the server.
 \item Apply circuit gates until other qubits are needed.
 \item (Parallel) Check if the server can execute any compatible public Clifford operation with the encrypted qubits it has. If so, the server is instructed to perform it and the client updates the Pauli frame.
 \end{enumerate}
 \item As long as there are still circuit operations to be executed:
 \begin{enumerate}
 \item The client determines which qubits are needed for the next operations.
 \item The client freshly encrypts the qubits it does not need, sends them to the server, and requests the qubits it needs from the server.
 \item The client decrypts the qubits it receives with the key $(x_j,z_j)$ of the corresponding qubit $j$ and the appropriate Pauli-frame update rules for the operations applied while the qubit was stored by the server.
 \item Apply circuit gates until other qubits are needed.
 \item If a qubit completes its last circuit operation, it is measured (if measurement is required). If it is measured by the server, the server communicates the result to the client and the client decrypts it. If no measurement is required, it is sent encrypted to the server so as not to take up space on the client.
 \item (Parallel) Check if the server can execute any compatible public Clifford operation with the encrypted qubits it has. If so, the server is instructed to perform it and the client updates the Pauli frame.
 \end{enumerate}
 \end{enumerate}
\end{protocol}

\paragraph{Protocol \ref{Protocol: 1}: assumptions and guarantees.}
$ $

\textbf{Parties.} One client and one server.

\textbf{Assumptions.} The client can store and operate on at most $M$ qubits. Every private local operation delegated back to the client has arity at most $M$, and the scheduler can bring the required qubits into the client's memory when that operation is due; otherwise that operation must be decomposed or implemented with another primitive. The same condition applies to non-Clifford gates: they are not evaluated homomorphically by the server in the base protocol, but are local client operations whenever their arity is at most $M$. The client generates fresh secret Pauli keys, and either measures locally or uses delegated readout only under the measurement restrictions stated below. Whenever the server acts on QOTP-encrypted data, the delegated public gate set is restricted to Clifford operations unless an additional non-Clifford encrypted-evaluation mechanism is specified, as in Assumption \ref{ass:clifford-only}.

\textbf{Leakage.} The server learns at least $N$, $M$, the number of communication rounds, the timing of the requests, which qubits are requested or returned, and the public gate instructions that it is asked to execute.

\textbf{Correctness.} If the client applies each private gate locally, executes each non-Clifford gate of arity at most $M$ by the local decrypt-apply-freshly-encrypt sequence, and updates the Pauli frame after each delegated public Clifford gate according to Proposition \ref{prop:clifford-update}, then decryption recovers the same state and measurement statistics as the target circuit.

\textbf{Privacy.} By Lemma \ref{lem:qotp}, each encrypted register stored by the server is information-theoretically hidden at the level of the server's reduced density matrix while the corresponding QOTP keys remain uniform and secret.

\textbf{Limitations.} The protocol does not by itself hide the public circuit layer, the communication pattern, or the timing information. The server still learns the communication pattern, timing and public Clifford skeleton unless these are hidden by the structural-privacy compiler. Non-Clifford operations of arity at most $M$ are supported by local execution rather than by server-side encrypted non-Clifford evaluation, at the price of extra communication, local computation and fresh QOTP encryption. Larger non-Clifford gates require decomposition or an additional primitive.

It is also useful to note that, even if the client is assumed not to be able to measure qubits, delegated readout remains straightforward only for Pauli-basis measurements. In particular, in the computational basis the server can measure the encrypted state and return a bit $b$, after which the client outputs $b\oplus x$; the hidden $Z$ key does not affect that outcome. More generally, measuring $X^xZ^z\rho Z^zX^x$ in a basis $\mathcal{B}$ is equivalent to measuring $\rho$ in the conjugated basis $Z^zX^x\mathcal{B}X^xZ^z$. Therefore, for arbitrary measurement bases the client must either adapt the requested basis to the hidden Pauli key or invoke an explicit measurement gadget or subprotocol. For this reason, the cases analyzed below do not treat the inability to measure as an isolated case, although the privacy statement remains the one from Lemma \ref{lem:qotp}: it concerns the encrypted premeasurement state and requires the Pauli keys to stay secret.

\begin{table}[t]
\centering
\caption{Pauli-basis readout corrections under the single-qubit QOTP convention $P_{x,z}=X^xZ^z$. Here $y_B$ is the raw server outcome when measuring in basis $B$, and $m_B$ is the corrected logical outcome.}
\label{tab:pauli-readout-corrections}
\begin{tabular}{ccc}
\toprule
Basis $B$ & Raw outcome & Corrected logical outcome \\
\midrule
$Z$ & $y_Z$ & $m_Z=y_Z\oplus x$ \\
$X$ & $y_X$ & $m_X=y_X\oplus z$ \\
$Y$ & $y_Y$ & $m_Y=y_Y\oplus x\oplus z$ \\
\bottomrule
\end{tabular}
\end{table}
The $Y$ row depends on the convention used to label the two $Y$ eigenstates. Once a convention is fixed, the Pauli-frame dependence is $x\oplus z$, up to a global inversion of the labels.

\subsection{\texorpdfstring{Client with $M$ independent single-qubit devices}{Client with M independent single-qubit devices}}\label{sec: M qubit}
In this case, we further restrict the capabilities of the client quantum computer: instead of being an $M$-qubit quantum computer, it is a set of $M$ independent single-qubit devices without the ability to perform multi-qubit operations.

Protocol \ref{Protocol: 1} must therefore be modified so that only public multi-qubit Clifford operations are performed by the server, while private and non-Clifford content is isolated into local single-qubit layers whenever the base protocol is used. At first glance, this could compromise privacy in the multi-qubit operations. The risk comes either from a private parameter contained in the gate, for example the angles of the $R_{ZZ}$ gates, or from the structure of the target operation, for example the shape of a Grover oracle. The first issue can often be avoided because, as noted above, these gates can be decomposed into a gate sequence such that the private angle is isolated in one or several single-qubit gates, which can be executed on the client computer. Thus the target circuit must admit an alternating public-Clifford/local-single-qubit decomposition of the form stated in Sec. \ref{sec: techniques}; multi-qubit non-Clifford gates are allowed only when their non-Clifford content can be compiled into the client's local single-qubit gates and their entangling part remains public Clifford.

The second issue requires redundant multi-qubit gates inside the circuit. That is, we add extra gates that, when applied together, form the identity, for example two $\CNOT$ gates in a row, and that are intended to be indistinguishable from original circuit gates from the server's point of view. However, this hiding can be defeated if the server optimizes the circuit it has received so as to cancel these gate pairs. To avoid this, we add requests for server qubits after each of the server operations (padding or private). The server should then no longer be able to tell from the interaction transcript which requests correspond to structural padding and which correspond to genuine private operations. An example is to apply $\CNOT$ on the server, request one of the qubits and apply a gate ($R_Z$ or Identity), and send it back to apply another $\CNOT$. The server then sees a pattern that is compatible with several private explanations.

However, if single-qubit gates are not applied in most of the circuit, it is impractical to hide the padding gates in this way. To deal with such cases, we can request those qubits and simply apply a routing permutation between the single-qubit devices, so that when the client resends them the mapping between logical wires and physical ports is randomized. In this setting the operation is a port permutation or wire relabeling, not a physical two-qubit $\SWAP$ gate; it updates which physical output port carries each logical qubit. To reduce side-channel leakage about which ports are disconnected, for example by sending light pulses or other signals, or by the associated delay times, all the client ports should be disconnected and reconnected at the same time. We consider both these routing permutations and these intermediate gates to be private operations in the structural leakage model introduced next.

Another alternative is to add several auxiliary qubits in states $\ket{0}$ or $\ket{1}$ to implement decoy gates, which in this case do not have to cancel each other. The $\ket{0}$ qubits control the decoy gates, while the $\ket{1}$ qubits control the true gates. However, this method is not safe for two reasons. First, if it is used for gates that contain private information in their angle, we still reveal that information to the server, even if we provide additional misleading angles. Second, for structural hiding it is not useful either, since these qubits are always $Z$-basis eigenstates, so they can be measured without altering the system, allowing the server to measure them without us noticing. Thus, the server can obtain the circuit if it can somehow determine what these auxiliary qubits are.

The structural-privacy claims in this protocol use the server-view criterion of Sec. \ref{sec: techniques}, in particular Criterion \ref{prop:structural-privacy}.

Therefore, after making these changes, the protocol is as follows.
\begin{protocol}\label{Protocol: 2}

 \textbf{Input}: Quantum circuit, private/public operations list, $N$ (qubits needed) and $M$ (qubits of the client).

 \textbf{Output}: Result of the circuit execution.
 \begin{enumerate}
 \item Let $Q=\lceil N/M\rceil$ be the number of client batches required to cover the $N$ logical qubits with an $M$-qubit client register.
 \item For $i=0,\ldots,Q-2$:
 \begin{enumerate}
 \item Generate a group of $M$ qubits, from qubit $iM$ to qubit $(i+1)M-1$.
 \item Apply circuit single-qubit gates until multi-qubit gates are needed.
 \item Encrypt the qubits by applying $X^{x_j}Z^{z_j}$ to qubit $j$, with fresh random keys $x_j,z_j\in \{0,1\}$.
 \item Send the group of qubits to the server.
 \item (Parallel) Check if the server can execute any compatible public Clifford operation with the encrypted qubits it has. If so, the server is instructed to perform it and the client updates the Pauli frame.
 \end{enumerate}
 \item Last round:
 \begin{enumerate}
 \item Generate the last qubits, from qubit $(Q-1)M$ to qubit $N-1$.
 \item If the client currently holds $M'<M$ qubits, it asks the server for $M-M'$ qubits that must undergo private operations with those held by the client. The client decrypts the qubits it receives with the key $(x_j,z_j)$ of the corresponding qubit $j$ and the appropriate Pauli-frame update rules for the operations applied while the qubit was stored by the server.
 \item Apply circuit single-qubit gates until multi-qubit gates are needed.
 \item (Parallel) Check if the server can execute any compatible public Clifford operation with the encrypted qubits it has. If so, the server is instructed to perform it and the client updates the Pauli frame.
 \end{enumerate}
 \item As long as there are still circuit operations to be executed:
 \begin{enumerate}
 \item The client determines which qubits are needed for the next operations.
 \item The client applies a random logical-to-physical routing permutation to the port assignment for the qubits it does not need, encrypts them with fresh keys, sends them to the server, and requests the qubits it needs.
 \item The client decrypts the qubits it receives with the key $(x_j,z_j)$ of the corresponding qubit $j$ and the appropriate Pauli-frame update rules for the operations applied while the qubit was stored by the server.
 \item Apply circuit single-qubit gates until multi-qubit gates are needed.
 \item If a qubit completes its last circuit operation, it is measured. If it is measured by the server, the server communicates the result to the client and the client decrypts it.
 \item (Parallel) Check if the server can execute any compatible public Clifford operation with the encrypted qubits it has. If so, the server is instructed to perform it and the client updates the Pauli frame.
 \end{enumerate}
 \end{enumerate}
\end{protocol}

\paragraph{Protocol \ref{Protocol: 2}: assumptions and guarantees.}
$ $

\textbf{Parties.} One client and one server.

\textbf{Assumptions.} The client can perform universal single-qubit gates in this base Protocol \ref{Protocol: 2} setting, local port routing between its single-qubit devices, and fresh QOTP encryption. The delegated public gate set acting on encrypted data is again restricted by Assumption \ref{ass:clifford-only}. The admissible circuits are those admitting the alternating decomposition $U_{\mathrm{target}}=C_LD_LC_{L-1}D_{L-1}\cdots C_1D_1C_0$, where the server implements each public Clifford layer $C_j$ and the client implements each local single-qubit layer $D_j$. The port randomization is assumed to hide which logical wire is attached to which physical output port, except for the declared leakage. Any structural-privacy claim also assumes a randomized compiler whose output distributions satisfy Criterion \ref{prop:structural-privacy} for the circuit components being hidden.

\textbf{Leakage.} The server learns at least $N$, $M$, the number of rounds, the public multi-qubit gates it is asked to apply, and any residual information leaked by the port hardware, timing and measurement pattern.

\textbf{Correctness.} The protocol is correct for the same reason as Protocol \ref{Protocol: 1}, for circuits compiled into the allowed public-Clifford/local-single-qubit layers, with the additional observation that the inserted padding gates cancel and the routing permutations merely relabel qubits at the logical level.

\textbf{Privacy.} Qubit contents remain hidden by Lemma \ref{lem:qotp} and Theorem \ref{thm:protocol1-2-qotp-state-privacy}, while structural privacy is obtained only to the extent that the resulting distribution of $\mathsf{View}$ depends on the hidden circuit component through the declared leakage $\mathcal{L}$, as formalized in Criterion \ref{prop:structural-privacy}. This is a leakage-dependent guarantee for the chosen randomized compilation procedure, not a generic guarantee for arbitrary circuit transformations.

\textbf{Limitations.} This protocol does not establish unconditional hiding of the circuit structure against arbitrary side channels, does not allow the server to apply non-Clifford gates directly to encrypted data, and supports non-Clifford behavior only through the alternating decomposition above or through an explicit extra primitive. The decomposition also increases gate count, communication rounds and re-encryption steps.

\begin{theorem}[State privacy of Protocols~\ref{Protocol: 1} and~\ref{Protocol: 2} under fresh QOTP keys]\label{thm:protocol1-2-qotp-state-privacy}
Consider any $n$-qubit delegated register in Protocol~\ref{Protocol: 1} or Protocol~\ref{Protocol: 2}. Suppose the client samples fresh, uniform and secret QOTP keys $x,z\in\{0,1\}^n$ for that register, independently of the plaintext state and of the declared leakage $\mathcal{L}$. In the strict honest-but-curious model, and also in the purified honest-but-curious model for ancillas compatible with the prescribed public Clifford implementation, the server's quantum marginal for the encrypted delegated register is maximally mixed, conditioned on $\mathcal{L}$. The remaining information is therefore the declared leakage, such as register size, timing, routing leakage and public gate pattern. Protocol~\ref{Protocol: 2} additionally needs the separate structural-privacy compiler assumptions stated below; those assumptions are not part of the QOTP state-privacy claim itself.
\end{theorem}
\begin{proof}[Proof]
For any plaintext register $\rho$ and $P_{x,z}=X^xZ^z$ in the tensor-product convention of Sec.~\ref{sec: techniques},
\[
\mathbb{E}_{x,z}\!\left[P_{x,z}\rho P_{x,z}^{\dagger}\right]=\frac{I}{2^n}.
\]
During delegated public Clifford layers, Proposition~\ref{prop:clifford-update} gives $CP_{x,z}=\omega P_{f_C(x,z)}C$, so the operation only changes the secret Pauli frame known to the client. Hence the register seen by the server remains Pauli-twirled during the Clifford part of the delegation. This theorem is a state-privacy statement for the encrypted marginal; it does not hide the metadata explicitly included in $\mathcal{L}$.
\end{proof}

\section{\texorpdfstring{Protocols for resource-limited clients}{Protocols for resource-limited clients}}\label{sec: protocols user}
Now consider a situation in which the client is a private user. Here, the user has few quantum computational resources, perhaps only one or two qubits, either on a personal device or through access to a trusted local device.

\subsection{\texorpdfstring{Small-client special cases}{Small-client special cases}}
The smallest quantum-client settings are best viewed as boundary cases of Protocols \ref{Protocol: 1} and \ref{Protocol: 2}, rather than as separate protocols.

\paragraph{\texorpdfstring{Two-qubit client.}{Two-qubit client.}}
A fully capable two-qubit client instantiates Protocol \ref{Protocol: 1} with $M=2$: it can execute any private one- or two-qubit operation locally, but larger private or non-Clifford blocks still require decomposition or another primitive. Key generation must be scheduled around the computation because neither qubit can be permanently reserved for randomness.

\paragraph{\texorpdfstring{Two independent single-qubit devices.}{Two independent single-qubit devices.}}
Two independent one-qubit devices instantiate Protocol \ref{Protocol: 2} with $M=2$: the client can prepare, encrypt, route and apply local single-qubit gates on two qubits in parallel, while entangling Clifford gates remain delegated to the server. 

\paragraph{\texorpdfstring{Single-qubit client.}{Single-qubit client.}}
A single-qubit client instantiates the QOTP state-privacy component of Protocol \ref{Protocol: 2} with $M=1$, but it cannot use local port permutations to hide structure and cannot dedicate a separate qubit to key generation. 

\begin{remark}[\texorpdfstring{Structural hiding degenerates for one device}{Structural hiding degenerates for one device}]\label{rem:single-device-structure}
The QOTP state-privacy component of Protocol \ref{Protocol: 2} still applies when $M=1$, provided the client can generate fresh Pauli keys and exchange the encrypted qubit with the server.
However, the structural-hiding mechanism based on local routing permutations degenerates: with only one client qubit, there is no non-trivial local permutation of ports. Therefore, in cases where circuit structure must be hidden, the protocol must rely on other gates hidden from the server, on an MBQC method, or on another hiding primitive, with the associated increase in server resources. One can also combine the Sec. \ref{sec: 0 qubit} technique with the client qubit to deal with the routing problem by making it an additional routing node.
\end{remark}

\subsection{\texorpdfstring{Restricted single-qubit devices and routing-hidden Pauli-axis rotation sharing}{Restricted single-qubit devices and routing-hidden Pauli-axis rotation sharing}}\label{sec:restricted-rz}

Consider $M$ restricted single-qubit devices that cannot perform all the desired operations. Suppose they can only perform $X$ and $Z$, the minimum needed for the QOTP masks. In this case, although the structure of the circuit can be hidden to the extent allowed by routing permutations, a single-qubit gate angle cannot be hidden by applying it locally, since the client cannot perform it. Consequently, the local non-Clifford strategy of Protocol \ref{Protocol: 2} is available only when every required local gate $d_{j,i}$ belongs to the restricted gate set. If the device can only apply $X$ and $Z$, then generic $T$, $S$, $R_Z(\theta)$, $R_X(\theta)$, $R_Y(\theta)$ or arbitrary single-qubit rotations are not available unless an additional trusted local module is supplied. For private one-qubit Pauli-axis rotations $R_X$, $R_Y$ and $R_Z$ that the restricted client cannot apply locally, this section gives the reusable finite-grid routing-hidden sign-randomized $r$-share angle-ambiguity primitive used in Protocol \ref{Protocol: 3}. Its privacy is leakage-dependent angle ambiguity under hidden signs, hidden grouping, hidden routing, finite-grid masking and side-channel assumptions; it should not be read as full blindness or as generic non-Clifford homomorphic evaluation.

\paragraph{\texorpdfstring{Finite angular grid.}{Finite angular grid.}}
Fix a public precision parameter $p$ and let $q=2^p$. Define
\begin{equation}
 \Lambda_p
 =
 \left\{
 \frac{2\pi k}{q}:k\in\mathbb{Z}_q
 \right\}.
\end{equation}
Angles used by the primitive are represented by labels in $\mathbb{Z}_q$, and the physical rotation associated with $a\in\mathbb{Z}_q$ is $R_A(2\pi a/q)$. All angle shares, masks and signed sums in the primitive are computed modulo $q$. The parameter $p$, the grid $\Lambda_p$ and the rounding or compilation rule that maps a requested physical angle to a grid label are part of the implementation model and of the public leakage. The private value is the target grid label, not the existence of the grid.

\begin{remark}[Finite-precision implementation note]\label{rem:finite-precision-angle-sharing}
The finite grid
\[
\Lambda_p=\left\{2\pi k/2^p : k\in\mathbb{Z}_{2^p}\right\}
\]
should be understood as the operational angle domain for the angle-sharing primitives. Additions are performed modulo $2^p$ at the label level. This avoids relying on ideal continuous randomness and makes the precision parameter, rounding rule and leakage model explicit.
\end{remark}

\begin{lemma}[Finite-grid two-share sanity check]\label{lem:angle-sharing}
Let the target label be $\theta\in\mathbb{Z}_q$, sample $\theta_1$ uniformly from $\mathbb{Z}_q$, and define
\begin{equation}
 \theta_2=\theta-\theta_1 \pmod{q}.
\end{equation}
Then $\theta_1$ and $\theta_2$ are each individually uniform on $\mathbb{Z}_q$ and therefore reveal no information about $\theta$ on their own. Moreover,
\begin{equation}
 R_Z(2\pi\theta_2/q)R_Z(2\pi\theta_1/q)=R_Z(2\pi\theta/q),
\end{equation}
up to the irrelevant global phase associated with the convention that $R_Z(2\pi a/q)$ denotes the finite-grid rotation associated with label $a$ and equality is modulo $q$.
\end{lemma}

Lemma \ref{lem:angle-sharing} is only a finite-grid algebraic baseline and is not the primitive used in Protocol \ref{Protocol: 3}. If the same adversarial server sees all shares associated with one logical rotation and can group them correctly, then ordinary additive sharing lets it recover the signed sum modulo $q$. The strengthened construction below therefore combines finite-grid $r$-share masking with hidden signs, routing-hidden grouping, masked idle rotations and fresh per-shot randomness. Its purpose is to make it difficult for the server to recover both the hidden signs of the shares and the grouping of shares belonging to the same logical rotation.

\begin{lemma}[\texorpdfstring{QOTP sign action on $R_Z$ rotations}{QOTP sign action on R_Z rotations}]\label{lem:qotp-rz-sign}
Let $P_{x,z}=X^xZ^z$ be the single-qubit QOTP mask, with $x,z\in\{0,1\}$ unknown to the server. In the angle-ambiguity primitive, the state is already QOTP-encrypted before the first visible share is applied. If the server applies a visible finite-grid rotation $R_Z(\alpha)$, then
\begin{equation}
 R_Z(\alpha)X^xZ^z\ket{\psi}
 =
 X^xZ^zR_Z\!\left((-1)^x\alpha\right)\ket{\psi}.
\end{equation}
Equivalently, up to the same Pauli mask $P_{x,z}$, the logical state evolves as if $R_Z((-1)^x\alpha)$ had been applied to $\ket{\psi}$.
\end{lemma}

The identity follows from
\begin{equation}
 R_Z(\alpha)X = X R_Z(-\alpha), \qquad R_Z(\alpha)Z = ZR_Z(\alpha).
\end{equation}
Thus a hidden $X$ key flips the effective sign of a visible $R_Z$ angle, whereas a hidden $Z$ key does not change that angle in a computational-basis phase rotation.

For clarity, QOTP contributes in two distinct ways. Its usual role is state privacy: without the Pauli keys, the server's quantum marginal is Pauli-twirled. In the angle-sharing primitive, the hidden Pauli frame also induces hidden signs, or more generally Pauli-frame-dependent interpretations, for Pauli-axis rotations. State privacy and angle ambiguity should therefore not be conflated: the latter depends on the combination of QOTP-hidden signs, finite-grid $r$-share decomposition, routing and hidden grouping assumptions.

\begin{proposition}[\texorpdfstring{Finite-grid routing-hidden sign-randomized $R_Z$ sharing}{Finite-grid routing-hidden sign-randomized R_Z sharing}]\label{prop:sign-hidden-rz-blocks}
For a target private rotation $R_Z(2\pi\theta/q)$, first compile the physical angle to a target label $\theta\in\mathbb{Z}_q$ under the public grid model above, and choose a share parameter $r\geq 2$. The value of $r$ should be treated as a tunable security/overhead parameter, not as a fixed guarantee. The client samples hidden signs
\begin{equation}
s_i\in\{-1,+1\},\qquad i=1,\ldots,r,
\end{equation}
independently and uniformly. These signs are implemented by setting the effective hidden $X$ key before the $i$-th visible rotation so that
\begin{equation}
 s_i=(-1)^{x_i}.
\end{equation}
The client then samples uniform grid labels
\begin{equation}
 \alpha_1,\ldots,\alpha_{r-1}\leftarrow\mathbb{Z}_q
\end{equation}
and defines the final share by
\begin{equation}
 \alpha_r
 =
 s_r\left(
 \theta-\sum_{i=1}^{r-1}s_i\alpha_i
 \right)
 \pmod{q}.
\end{equation}
Then
\begin{equation}
 \sum_{i=1}^r s_i\alpha_i=\theta \pmod{q},
\end{equation}
and therefore
\begin{equation}
 \prod_{i=1}^r R_Z(2\pi s_i\alpha_i/q)=R_Z(2\pi\theta/q),
\end{equation}
up to an irrelevant global phase and under the finite-grid convention. By Lemma \ref{lem:qotp-rz-sign}, the server can be instructed to apply the visible rotations $R_Z(2\pi\alpha_i/q)$ while the hidden $X$ keys determine the signed logical rotations $R_Z(2\pi s_i\alpha_i/q)$. Thus the visible share labels are not the logical rotation labels by themselves: their logical signs are induced by the hidden Pauli frame on a QOTP-encrypted state.
\end{proposition}

Equivalently, in physical-angle notation, the algebraic correctness statement is
\[
\sum_{i=1}^r s_i\alpha_i=\theta \pmod{2\pi}
\quad\Longrightarrow\quad
\prod_{i=1}^r R_Z(s_i\alpha_i)=R_Z(\theta),
\]
up to global phase. If the grouping of the $r$ shares is known but the signs are hidden, the visible transcript is compatible with
\[
\Theta(\alpha)=
\left\{
\sum_{i=1}^r t_i\alpha_i \pmod{2\pi}: t_i\in\{-1,+1\}
\right\},
\]
whose size is at most $2^r$. Hidden grouping can increase the matching ambiguity across transcripts, but this ambiguity statement is not, by itself, a simulator-based blindness theorem. Full simulator-based angle blindness requires a separate indistinguishability proof for the chosen compiler, finite precision model and leakage function.

\paragraph{\texorpdfstring{Fresh sign requirement.}{Fresh sign requirement.}}
The independence of the signs requires that the effective $X$ key before each share be freshly sampled or deliberately reset by the client. Operationally, this means that between share rounds the qubit must pass through the client, or through a trusted local re-encryption step, so that the client can impose the next hidden sign without revealing it to the server. If the same $X$ key is reused across all shares, then all visible rotations acquire the same sign and the $2^r$ sign ambiguity does not hold.

If the server can group all $r$ shares of a logical rotation, then the signs alone leave at most the finite-grid candidate set
\begin{equation}
 \Theta_{p}(\alpha)
 =
 \left\{
 \sum_{i=1}^r t_i\alpha_i
 \pmod{q}
 :
 t_i\in\{-1,+1\}
 \right\}
 \subseteq\mathbb{Z}_q.
\end{equation}
Equivalently, $\Theta_p(\alpha)\subseteq\mathbb{Z}_q$, and
\begin{equation}
 |\Theta_p(\alpha)|\leq \min\{2^r,q\}.
\end{equation}
This is only an ambiguity set, not full blindness, and the bound can be strict because of finite-grid collisions, structured priors, repeated angles, parameters generated by optimizers, correlations between shots or side-channel leakage. Small $r$, including $r=2$, may give practical ambiguity when the server's goal is to infer an entire vector of many angles and the grouping remains hidden.

The share parameter $r$ should be chosen relative to an informal target ambiguity level $\lambda$. In the idealized case where signed sums are distinct and no side-channel or grouping information is leaked, a single $r$-share target gives at most $\min\{r,p\}$ bits of sign-sum ambiguity, since the number of possible signed sums is bounded by both $2^r$ and the grid size $q=2^p$. For a masked two-target construction, the idealized net-angle ambiguity is bounded by $\min\{2r,p\}$ bits. These are upper bounds on transcript ambiguity, not composable security guarantees; finite-grid collisions, repeated parameters, optimizer correlations, priors and cross-shot alignment can reduce the effective ambiguity.

\paragraph{\texorpdfstring{Routing-hidden sign-randomized $R_Z$ layer.}{Routing-hidden sign-randomized R_Z layer.}}
Now consider a layer with $N$ visible qubits. Let $n$ be the number of actual logical $R_Z$ rotations in the layer, and let the remaining visible qubits be idle qubits of the real circuit, not necessarily extra trap qubits. The server should see $R_Z$ instructions on many or all visible qubits in each share round. Between share rounds, the client applies hidden routing permutations or port randomization so that the server cannot reliably match the share applied to one physical port in round $i$ with the corresponding logical qubit in round $i+1$. In an idealized layer with $N$ visible rotations per share round and no side-channel information, the number of possible matchings between $r$ rounds scales like
\begin{equation}
 (N!)^{r-1}.
\end{equation}
This is an idealized transcript-ambiguity estimate, not a composable security proof.

Directly sharing an idle target as $R_Z(0)$ can create exact algebraic signatures. In particular, for $r=2$, zero targets can produce a visible relation of the form $\alpha_2=\pm\alpha_1$, which may help the server identify dummy groups or recover the routing. Using $r\geq 3$ reduces simple algebraic signatures in naive zero-target sharing, but the masked dummy construction is preferable because it avoids making zero targets special in the visible transcript.

For each visible qubit $j$ in the layer, define
\begin{equation}
 \theta_j=
 \begin{cases}
 \text{the intended private grid label}, & \text{if qubit $j$ receives a real $R_Z$ rotation},\\
 0, & \text{if qubit $j$ is idle in this layer}.
 \end{cases}
\end{equation}
Sample a fresh angular mask uniformly
\begin{equation}
 \eta_j\leftarrow\mathbb{Z}_q
\end{equation}
and replace $R_Z(2\pi\theta_j/q)$ by the logically equivalent pair
\begin{equation}
 R_Z(-2\pi\eta_j/q)R_Z(2\pi\theta_j/q+2\pi\eta_j/q)=R_Z(2\pi\theta_j/q),
\end{equation}
with all labels computed modulo $q$. For idle qubits this becomes
\begin{equation}
 R_Z(-2\pi\eta_j/q)R_Z(2\pi\eta_j/q)=I,
\end{equation}
up to global phase, while both intermediate target grid labels are marginally uniform and do not reveal that the net operation is the identity. Define
\begin{equation}
 \vartheta_{j,0}=\theta_j+\eta_j \pmod{q},
 \qquad
 \vartheta_{j,1}=-\eta_j \pmod{q}.
\end{equation}
Each target rotation $\vartheta_{j,b}$ is implemented using the sign-randomized $r$-share construction, with hidden signs $s_{j,b,i}$ and visible shares $\alpha_{j,b,i}$ satisfying
\begin{equation}
 \sum_{i=1}^r s_{j,b,i}\alpha_{j,b,i}
 =
 \vartheta_{j,b}
 \pmod{q}.
\end{equation}
The layer construction is:
\begin{enumerate}
 \item Input a set of $N$ visible qubits, intended logical grid labels $\theta_j$ with $\theta_j=0$ for idle qubits, a share parameter $r$, the public precision parameter $p$, and hidden routing permutations between share rounds.
 \item For each qubit $j$, sample $\eta_j\leftarrow\mathbb{Z}_q$.
 \item Set $\vartheta_{j,0}=\theta_j+\eta_j \pmod{q}$ and $\vartheta_{j,1}=-\eta_j \pmod{q}$.
 \item For each $b\in\{0,1\}$, sample $s_{j,b,i}\in\{-1,+1\}$ uniformly and independently, sample $r-1$ shares uniformly from $\mathbb{Z}_q$, and close the last share so that $\sum_i s_{j,b,i}\alpha_{j,b,i}=\vartheta_{j,b}\pmod{q}$.
 \item Before each visible share, the client sets the effective hidden $X$ key so that the sign seen logically is $s_{j,b,i}$.
 \item The server applies the visible $R_Z(2\pi\alpha_{j,b,i}/q)$ instructions to the physical ports it sees.
 \item Between share rounds, the client applies hidden routing permutations or port randomization.
\end{enumerate}
Correctness follows because, with the products ordered according to the circuit time order,
\begin{equation}
 \prod_{b=0}^1\prod_{i=1}^r
 R_Z(2\pi s_{j,b,i}\alpha_{j,b,i}/q)
 =
 R_Z(2\pi\vartheta_{j,1}/q)R_Z(2\pi\vartheta_{j,0}/q)
 =
 R_Z(2\pi\theta_j/q).
\end{equation}
For idle qubits, where $\theta_j=0$, the net operation is the identity. This avoids adding extra physical qubits when the dummy positions are idle qubits already present in the circuit. Extra verifier-circuit qubits from the trap-based detection layer may also be used as decoys if available, but they are not required by the angle-ambiguity construction.

The decoy positions in this angle-ambiguity construction need not be additional trap qubits. They may be idle qubits of the real circuit in that layer. If verifier-circuit qubits from the trap-based detection layer are also used as decoys, this must be accounted for in both the verification and leakage analyses.

When this primitive is used inside the padded QAOA/QUBO compiler of Sec. \ref{sec: techniques}, the two logical targets $-\eta$ and $\theta+\eta$ are not exposed as a visible pair. They are independently decomposed, sign-randomized, routed, shuffled and mixed with other visible slots of the same operation class. If a server could identify both target groups and their effective signs, then the net angle could reveal $\theta$. The security claim therefore excludes that information through the hidden-sign and hidden-grouping assumptions and through the side-channel model.

\paragraph{Leakage and ambiguity.}
The construction is not, by itself, a simulator-based or composable blindness theorem and is not claimed to give unconditional angle blindness. It provides leakage-dependent angle ambiguity under hidden signs, hidden grouping, hidden routing, finite-grid masking and side-channel assumptions:
\begin{itemize}
 \item the signs are fresh, independent and hidden from the server;
 \item the effective $X$ keys used to generate those signs are not leaked;
 \item routing permutations or port randomization hide the grouping of shares across rounds;
 \item side channels such as timing, physical port labels, repeated ordering, cross-shot alignment, loss patterns, deterministic batching, optimizer traces, visible classical key-update patterns, pulse shape or routing latency do not reveal the logical matching;
 \item dummy and idle rotations are distributionally indistinguishable from real rotations because of the finite-grid angular masks $\eta_j$;
 \item fresh randomness is used independently in every shot;
 \item cross-shot correlations do not reveal the same logical share groups.
\end{itemize}
If grouping is known, the server can still form a finite candidate set of possible signed sums, namely $\Theta_p(\alpha)$ above. For $n$ independent real rotations, if all groupings are known but signs are hidden and all candidates are distinct, an idealized exact guess of all signs succeeds with probability $2^{-rn}$; in the masked two-target construction, the analogous estimate uses the corresponding sign ambiguity over the $2r$ signed shares. If grouping is also hidden, the ambiguity is further amplified by the hidden matching problem, ideally by factors such as $(N!)^{r-1}$ for $r$ rounds, or by a larger matching ambiguity when the masked two-target construction produces $2r$ share rounds. These estimates quantify transcript ambiguity under the stated leakage model; they do not replace a simulator-based or composable blindness proof. The finite-grid bound $|\Theta_p(\alpha)|\leq\min\{2^r,q\}$ can be smaller in concrete transcripts because of grid collisions, structured priors, repeated angles, optimizer-generated parameter correlations, cross-shot correlations, visible key-update regularities or side-channel leakage.

\paragraph{\texorpdfstring{Fresh randomness per shot.}{Fresh randomness per shot.}}
For repeated shots of the same circuit, all randomness must be freshly sampled:
\begin{equation}
 s_{j,b,i}^{(\mathrm{shot})},
 \quad
 \alpha_{j,b,i}^{(\mathrm{shot})},
 \quad
 \eta_j^{(\mathrm{shot})},
 \quad
 \pi_i^{(\mathrm{shot})}
\end{equation}
are sampled independently for every shot. Fresh randomness per shot helps prevent the server from correlating candidate sets across repeated executions. However, this protection relies on the assumption that the server cannot align the same hidden logical rotation across shots using side channels or declared leakage. If the server can reliably identify corresponding logical share groups across shots, it may intersect candidate sets and reduce the ambiguity.

\paragraph{Overhead.}
The cost of the $Z$-axis instance of the strengthened construction is an increase in visible $R_Z$ operations, communication rounds, re-encryption steps and routing permutations. In the masked idle-rotation version, each logical $R_Z(\theta_j)$ or masked identity $I$ is represented by two masked target rotations, each split into $r$ sign-randomized shares, for a total of $2r$ visible $R_Z$ share applications per visible qubit in that layer. This avoids adding extra physical qubits when idle real-circuit qubits are available, but it increases circuit depth and communication overhead.

This observation is closely related to the problem of evaluating non-Clifford rotations on encrypted data. In general, asking the server to apply a one-qubit Pauli-axis rotation directly to a QOTP-encrypted qubit does not reduce to an ordinary public-operation Pauli-frame update, because the logical sign of the applied angle depends on the hidden Pauli mask that anticommutes with the rotation axis. That dependence can be used as part of the dedicated angle-ambiguity construction in Proposition \ref{prop:sign-hidden-rz-blocks}, but it does not by itself establish privacy of the full transcript. The routing-hidden sign-randomized Pauli-axis rotation construction is therefore not the default mechanism for non-Clifford evaluation in the base protocols; it should be read as a separate leakage-dependent angle-ambiguity primitive under its stated leakage assumptions, not as a generic non-Clifford encrypted-evaluation capability. To implement arbitrary private non-Clifford operations on encrypted data one still needs adapted angles, gate teleportation, a quantum homomorphic or gadget-based primitive, local application by the client, or another explicit assumption.

The sign-randomization mechanism is not specific to $Z$ rotations. For a one-qubit QOTP mask $X^aZ^b$, Pauli-axis rotations obey, up to irrelevant global phases and with finite-grid labels,
\begin{equation}
 R_Z(\alpha)X^aZ^b=X^aZ^bR_Z((-1)^a\alpha),
\end{equation}
\begin{equation}
 R_X(\alpha)X^aZ^b=X^aZ^bR_X((-1)^b\alpha),
\end{equation}
\begin{equation}
 R_Y(\alpha)X^aZ^b=X^aZ^bR_Y((-1)^{a\oplus b}\alpha).
\end{equation}
Thus, the hidden sign is determined by the QOTP component that anticommutes with the rotation axis: the $X$ mask flips the sign of a $Z$ rotation, the $Z$ mask flips the sign of an $X$ rotation, and the parity $a\oplus b$ flips the sign of a $Y$ rotation. The server applies the visible rotation $R_A(\alpha)$, but the effective logical rotation is $R_A(s\alpha)$, where $s$ is fixed by the hidden Pauli-frame component that anticommutes with $A$. Equivalently, with a single random sign bit $b_{\mathrm{s}}$, the client may use $X^{b_{\mathrm{s}}}$ for $R_Z$, $Z^{b_{\mathrm{s}}}$ for $R_X$, and either $X^{b_{\mathrm{s}}}$ or $Z^{b_{\mathrm{s}}}$ for $R_Y$. The client should not use $X^{b_{\mathrm{s}}}Z^{b_{\mathrm{s}}}$ with the same bit as the sign-flipping mask for $R_Y$, because this is proportional to $Y^{b_{\mathrm{s}}}$ and therefore commutes with $R_Y$. This extension preserves the same scope as the $R_Z$ construction: it gives finite-grid sign-randomized angle hiding for one-qubit Pauli-axis rotations under the same hidden-sign, hidden-grouping and routing assumptions, not arbitrary non-Clifford blindness or generic homomorphic evaluation of all rotations.

In summary, finite-grid routing-hidden sign-randomized Pauli-axis rotation sharing is an angle-ambiguity amplification primitive: it combines hidden signs, hidden grouping and distributionally masked idle rotations, with guarantees that remain leakage-dependent. 

However, this routing-permutation-based grouping-hiding strategy cannot be used with only one client qubit, as noted in Remark \ref{rem:single-device-structure}. In that case, hiding private angles would require a different primitive, such as MBQC, gate teleportation, a homomorphic or gadget-based rotation mechanism, or a trusted or non-colluding multi-party implementation.

\begin{protocol}\label{Protocol: 3}
\textbf{Input}: A circuit compiled into public Clifford layers and private one-qubit Pauli-axis rotations, a restricted quantum client that can apply QOTP Paulis and routing permutations, the number of qubits $N$, the number $M$ of client qubits or restricted devices, the public precision parameter $p$ and the share parameter $r$.
 \begin{enumerate}
 \item Let $Q=\lceil N/M\rceil$ be the number of client batches required to cover the $N$ logical qubits with an $M$-qubit client register.
 \item The client prepares or receives the next qubits, applies fresh QOTP masks $X^{x_j}Z^{z_j}$, sends only encrypted qubits to the server, and keeps the Pauli keys and logical-to-physical routing map secret. Public Clifford operations are delegated as in Proposition \ref{prop:clifford-update}, with classical Pauli-frame updates by the client.
 \item Whenever a private Pauli-axis rotation $R_A(\theta)$, with $A\in\{X,Y,Z\}$, cannot be applied locally by the restricted client, the client does not send a bare additive sharing of $\theta$. Instead, the target angle is first compiled to a finite grid label $\theta\in\mathbb{Z}_q$, with $q=2^p$, and the rotation is implemented using the routing-hidden sign-randomized $r$-share angle-ambiguity primitive of Sec. \ref{sec:restricted-rz}, with Proposition \ref{prop:sign-hidden-rz-blocks} giving the $Z$-axis instance. The input qubit is QOTP-encrypted before the visible share sequence; the server receives only visible grid shares and physical ports; hidden Pauli-frame signs determine the logical signs of the visible rotations; and routing permutations or port randomization hide the grouping of shares between rounds.
 \item Before each visible share round, the client or trusted local re-encryption step refreshes the relevant hidden Pauli-frame component so that the desired sign is induced without revealing it to the server. Between share rounds, the client applies hidden routing permutations or port randomization, and all masks, shares and routing choices are freshly sampled per shot.
 \item The protocol repeats the public Clifford layers and the finite-grid angle-ambiguity blocks until the circuit is complete. If a qubit is measured by the server, the client interprets the raw outcome using the current QOTP keys; more general private measurements require the corresponding measurement subprotocol.
 \end{enumerate}
\end{protocol}

\paragraph{\texorpdfstring{Protocol \ref{Protocol: 3}: assumptions and guarantees.}{Protocol 3: assumptions and guarantees.}}
$ $

\textbf{Parties.} One restricted quantum client and one honest-but-curious quantum server.

\textbf{Assumptions.} The client can apply fresh QOTP masks, maintain hidden Pauli-frame keys, route or randomize physical ports between share rounds, and enforce fresh randomness for every shot. The server-side public operations on encrypted data remain restricted to the Clifford case unless another explicit non-Clifford primitive is supplied.

\textbf{Privacy.} QOTP gives state privacy for the encrypted quantum register as formalized in Proposition \ref{prop:protocol3-state-structural-privacy}. Structural privacy remains leakage-dependent through the routing and padding model. Private Pauli-axis rotations use finite-grid angle ambiguity under the hidden signs, hidden grouping, hidden routing, finite-grid masking and side-channel assumptions stated above. This angle-ambiguity primitive is not a full simulator-based or composable blindness theorem, and visible grid shares should not be interpreted as direct logical rotations.

\begin{proposition}[State and structural privacy scope of Protocol~\ref{Protocol: 3}]\label{prop:protocol3-state-structural-privacy}
Protocol~\ref{Protocol: 3} inherits the QOTP state privacy of Theorem~\ref{thm:protocol1-2-qotp-state-privacy} for each encrypted delegated register, provided fresh secret Pauli keys are used and the server follows the prescribed honest-but-curious public Clifford and share-application behavior. Structural privacy is obtained only relative to a specified randomized compiler whose server-view distributions are indistinguishable for circuits inside the same declared leakage class.
\end{proposition}
\begin{proof}[Proof sketch]
The state-privacy argument is identical to the Protocol~\ref{Protocol: 2} argument on every delegated batch: the server receives a Pauli-twirled marginal, and public Clifford gates only update the secret Pauli frame. The additional angle-sharing blocks do not by themselves hide the computation structure. Splitting the computation into client-side fragments, server-side Clifford fragments and visible rotation-share fragments may leak metadata unless the compiler pads, routes and randomizes so that all circuits with the same declared leakage induce the same distribution of $\mathsf{View}$. Thus structural privacy is a compiler-relative indistinguishability statement, not a consequence of QOTP state privacy alone.
This is a conditional reduction statement rather than a composable security proof: if the stated key-hiding, hidden-routing and leakage-indistinguishability conditions hold, then the QOTP or transcript-ambiguity argument gives the claimed component guarantee.
\end{proof}

\subsection{\texorpdfstring{Classical-client matching-hidden split-QOTP branch}{Classical-client matching-hidden split-QOTP branch}}\label{sec: 0 qubit}
The classical-client setting is a separate branch of the hierarchy, not merely the limit $M=0$ of the quantum-client protocols. Here, the protocol uses two computational servers and a common node that reroutes systems between them. The security model is coalition-based: allowed coalitions may pool their local transcripts, but no allowed coalition receives both the computational value-share information and the common-node matching information or, in the recommended variant, the hidden refresh-share information needed to reconstruct aligned Pauli keys or grouped angle shares. The common node performs the routing permutations of the communication, informs the client what the new qubit labels are, and must therefore be modeled explicitly as trusted or at least honest, outside the forbidden full coalition, with its side channels controlled. We assume that each computational server can perform the delegated operations assigned to it. In this setting, the client is still classical, so it can generate and store classical angle shares and classical Pauli keys even though it cannot touch the quantum state directly. Operations whose privacy is structural are exposed only through their public operation classes and are hidden only if the padding strategy satisfies the structural-privacy criterion for the declared leakage; this section does not claim client-side QOTP, but it does use a server-side split-QOTP whose key alignment is hidden by the common node.

Protocol \ref{Protocol: 4} does not rely on client-side QOTP, since the classical client never directly touches the quantum register. Instead, it maintains a persistent matching-hidden split-QOTP through the computational servers and the common node. The common node hides the matching between Pauli-key share tables, so aligned per-register Pauli keys are not reconstructed by any allowed coalition, conditioned on the declared leakage. Operationally, the classical client is trusted and either knows the current matching or receives enough private alignment information from the common node to compute the relevant aligned Pauli-frame or sign bit; this private alignment information is not revealed to $S_1$, $S_2$ or allowed coalitions that should not learn the matching. The common node is therefore not a passive detail of the communication network: it is part of the trust and leakage model.

\begin{assumption}[Coalition and matching model for Protocol \ref{Protocol: 4}]\label{ass:protocol4-coalitions}
The server-side parties are
\[
\mathcal P=\{S_1,S_2,R\},
\qquad
\mathcal C_{\mathrm{P4}}=\{C\subsetneq \mathcal P\},
\]
where $S_1$ and $S_2$ are computational servers and $R$ is the common node. The full coalition $\{S_1,S_2,R\}$ is excluded. The computational servers $S_1$ and $S_2$ may pool their computational value-share transcripts; this is not enough to reconstruct aligned keys or grouped angles if $R$ does not leak the hidden matching. A coalition $\{S_1,R\}$ lacks the value shares held only by $S_2$, and a coalition $\{S_2,R\}$ lacks the value shares held only by $S_1$. Only the full coalition $\{S_1,S_2,R\}$ obtains both value shares and matching information.

The common node's matching permutations must not be leaked through timing, port identities, physical labels, routing metadata, message sizes, loss patterns, deterministic ordering or any side channel beyond the declared leakage. If such a side channel reveals the hidden matching, then the corresponding state, angle, output or structural privacy claim must be weakened or the leaked information must be added to $L(D)$.
\end{assumption}

\paragraph{\texorpdfstring{Matching-hidden split-QOTP.}{Matching-hidden split-QOTP.}}
\noindent\textbf{Definition (\(\epsilon_{\mathrm{key}}\)-key-hiding matching compiler).} For every allowed coalition \(C\in\mathcal C_{\mathrm{P4}}\), the aligned Pauli key \(K_t\) protecting each server-visible register must satisfy
\[
\Delta\left(
\Pr[K_t \mid \mathsf{View}_C,L],
\mathrm{Unif}(\{0,1\}^{2n_t})
\right)
\leq \epsilon_{\mathrm{key}}.
\]
Protocol \ref{Protocol: 4} state privacy under persistent split-QOTP and \(\epsilon_{\mathrm{key}}\)-key hiding is conditioned on this property.

For a batch of $m$ registers, the client samples computational key-share tables
\[
\kappa_i^{(1)}
=
(x_i^{(1)},z_i^{(1)}),
\qquad
\kappa_j^{(2)}
=
(x_j^{(2)},z_j^{(2)}),
\]
and the common node stores a matching permutation,
\[
\sigma_R\in S_m.
\]
 In the recommended high-security common-node refresh variant, \(R\) also holds a hidden refresh-share table \(\{\kappa_i^{(R)}\}_{i=1}^m\), where \(\kappa_i^{(R)}=(x_i^{(R)},z_i^{(R)})\).
This matching is hidden from the computational servers and from any allowed coalition not containing $R$. Coalitions containing $R$ may know $\sigma_R$, but under the non-total-collusion assumption they do not also possess both computational key-share tables. Thus no allowed coalition obtains the triple consisting of the $S_1$ share table, the $S_2$ share table and the common-node matching.
In the refresh variant, no allowed coalition obtains the corresponding full reconstruction data consisting of both computational share tables together with the hidden refresh share.

Without the optional refresh share, the base two-share aligned key would use \(\kappa_i^{(1)}\oplus\kappa_{\sigma_R(i)}^{(2)}\). In the recommended high-security construction, the aligned Pauli key protecting register $i$ uses the additional common-node refresh share,
\[
\kappa_i
=
\kappa_i^{(1)}
\oplus
\kappa_{\sigma_R(i)}^{(2)}
\oplus
\kappa_i^{(R)}.
\]
Equivalently,
\[
x_i=x_i^{(1)}\oplus x_{\sigma_R(i)}^{(2)}\oplus x_i^{(R)},
\qquad
z_i=z_i^{(1)}\oplus z_{\sigma_R(i)}^{(2)}\oplus z_i^{(R)}.
\]
Pooling the two computational servers' lists does not reveal the aligned per-register keys unless the relevant hidden alignment and, in the refresh variant, the hidden refresh share \(\kappa_i^{(R)}\) are also known. At every protocol cut $t$, the physical register satisfies the persistent QOTP invariant
\[
\rho_{\mathrm{phys}}^{(t)}
=
P_{K_t}
\rho_{\mathrm{log}}^{(t)}
P_{K_t}^{\dagger}.
\]
Every server-visible quantum register is QOTP-protected at every protocol cut. Thus split-QOTP is maintained throughout the computation, not merely at final measurement.

The \(\epsilon_{\mathrm{key}}\)-key-hiding condition for an allowed coalition $C\in\mathcal C_{\mathrm{P4}}$ is the hypothesis that
\[
\Delta
\left(
\Pr[K_t\mid \mathsf{View}_C,L],
\mathrm{Unif}(\{0,1\}^{2n_t})
\right)
\leq
\epsilon_{\mathrm{key}}.
\]
Under this condition, the coalition's quantum view is independent of the logical encrypted state up to the declared leakage and $\epsilon_{\mathrm{key}}$.

\begin{remark}[Recommended common-node Pauli refresh]\label{rem:protocol4-common-node-refresh}
The cleanest/high-security instantiation of the Protocol \ref{Protocol: 4} key-hiding condition adds a common-node Pauli refresh
\[
\kappa_i
=
\kappa_i^{(1)}
\oplus
\kappa_{\sigma_R(i)}^{(2)}
\oplus
\kappa_i^{(R)}.
\]
If \(\kappa_i^{(R)}\) is uniformly random and hidden from the computational servers, then the aligned Pauli key is uniform against the coalition \(\{S_1,S_2\}\), independently of the entropy of the matching. Coalitions containing \(R\) still miss one computational share under the non-total-collusion model. This gives a direct sufficient construction of the \(\epsilon_{\mathrm{key}}=0\) key-hiding condition, up to implementation leakage.

\end{remark}

\noindent\textbf{Effective key entropy in Protocol \ref{Protocol: 4}.} For the coalition \(\{S_1,S_2\}\), in the absence of the refresh share, hidden matching would be the only missing information. The induced aligned key would be
\[
K_i=\kappa_i^{(1)}\oplus\kappa_{\sigma_R(i)}^{(2)}.
\]
If \(\sigma_R\) is trivial or leaked, the distribution collapses. If the second-share table is unbalanced or the batch is too small, the induced key distribution can be far from uniform. Therefore matching entropy is useful only insofar as the matching compiler satisfies the explicit \(\epsilon_{\mathrm{key}}\)-key-hiding condition. The common-node Pauli refresh is the recommended high-security way to make this condition direct.

\begin{protocol}\label{Protocol: 4}
\textbf{Input}: A circuit compiled into public Clifford layers and private Pauli-axis rotation slots, the number $N$ of logical qubits, a batch size $m$, a public finite grid $\Lambda_p$, a share parameter $r$, and the declared leakage function $L$.

The matching-hidden branch is non-degenerate only when the hidden matching has nontrivial entropy. In particular, the case \(m=1\) provides no matching uncertainty and must be protected by an additional hidden refresh share, such as the common-node Pauli refresh.

 \begin{enumerate}
 \item Server $S_1$ prepares or receives the $N$ quantum registers. In the classical-client branch, Protocol \ref{Protocol: 4} does not allow the classical client to inject an arbitrary private quantum input directly. If \(S_1\) prepares or receives an unencrypted quantum input before the split-QOTP invariant is established, that input is not protected from \(S_1\). The state-privacy claim applies only after the persistent matching-hidden split-QOTP invariant, including the refresh share in the recommended variant, has been established, or to computations whose initial quantum state is public and whose private information is encoded later through hidden parameters, structure, masks, or readout. For each active batch of $m$ registers, the client samples split Pauli-key share tables $\{\kappa_i^{(1)}\}_{i=1}^m$ and $\{\kappa_j^{(2)}\}_{j=1}^m$, and, in the recommended high-security variant, the hidden refresh table \(\{\kappa_i^{(R)}\}_{i=1}^m\), while the common node $R$ stores the matching $\sigma_R$. This matching is hidden from the computational servers and from any allowed coalition not containing $R$; coalitions containing \(R\) may know it, but they do not also hold both computational share tables under the non-total-collusion assumption. The computational servers apply their Pauli-mask shares through the scheduled handoffs, so the recommended high-security aligned key on register $i$ is $\kappa_i=\kappa_i^{(1)}\oplus\kappa_{\sigma_R(i)}^{(2)}\oplus\kappa_i^{(R)}$; the two-share form \(\kappa_i^{(1)}\oplus\kappa_{\sigma_R(i)}^{(2)}\) is only the base construction used when no refresh share is included.
 \item The protocol maintains the invariant $\rho_{\mathrm{phys}}^{(t)}=P_{K_t}\rho_{\mathrm{log}}^{(t)}P_{K_t}^{\dagger}$ at every server-visible cut. The client updates the classical split key tables and the common node updates hidden matchings as registers move between $S_1$, $S_2$ and $R$.
 \item Public Clifford gates are applied by the computational server currently holding the relevant encrypted registers. If a Clifford gate $G$ induces the Pauli-frame update $K\mapsto A_G(K)$, then each share is updated linearly and sharewise as
 \[
 \kappa^{(h)}\mapsto A_G(\kappa^{(h)}),
 \]
 with no party reconstructing the total Pauli frame.
 For example, for $\CNOT_{c,t}$ each computational server updates its own value-share table by $x_t^{(h)}\leftarrow x_c^{(h)}\oplus x_t^{(h)}$ and $z_c^{(h)}\leftarrow z_c^{(h)}\oplus z_t^{(h)}$, leaving the other two bits unchanged; for $\CZ_{1,2}$ it updates $z_1^{(h)}\leftarrow z_1^{(h)}\oplus x_2^{(h)}$ and $z_2^{(h)}\leftarrow z_2^{(h)}\oplus x_1^{(h)}$. When the common-node refresh share is present, its Pauli share is updated by the same linear Clifford-frame rule. The common node or routing layer maintains only the hidden alignment and matching metadata needed to interpret which shares are paired after routing. These linear updates preserve the invariant that only aligned shares reconstruct the effective Pauli key, so Protocol \ref{Protocol: 4} does not reconstruct the full frame during Clifford layers.
 \item For a private Pauli-axis rotation slot $g$ with axis $A\in\{X,Y,Z\}$ and logical angle $\theta_g$, the client uses the aligned hidden Pauli key to compute the QOTP-corrected angle
 \[
 \widetilde{\theta}_g
 =
 (-1)^{\chi_A(x_g,z_g)}
 \theta_g,
 \]
 where
 \[
 \chi_Z(x,z)=x,
 \qquad
 \chi_X(x,z)=z,
 \qquad
 \chi_Y(x,z)=x\oplus z.
 \]
 The angle-sharing layer shares $\widetilde{\theta}_g$, not the bare logical angle $\theta_g$.
 \item The corrected angles $\widetilde{\theta}_g$ are implemented by the finite-grid sign-randomized $r$-share primitive of Sec. \ref{sec:restricted-rz}. In Protocol \ref{Protocol: 4}, the common node supplies the additional hidden inter-round matching and shuffling across each visible operation class. Auxiliary slots, if used, are dummy, padding, cancelling, masked-idle, decoy or transcript-completion slots with no verifier semantics at this stage. The grid, $\eta_g$-masking convention, signed-share equations, auxiliary-slot equations and matching-ambiguity estimate are inherited from Sec. \ref{sec:restricted-rz}; the mask $\eta_g$ does not hide the angle by itself.
 \item The protocol repeats the public Clifford updates, hidden matching updates and shuffled private rotation layers until the circuit is complete. For computational-basis output, the final readout uses the current persistent matching-hidden split-QOTP $X$ frame; the split-mask readout proposition below is the final-measurement special case of the same persistent frame, not a separate state-privacy mechanism.
 \end{enumerate}

\end{protocol}

\begin{proposition}[\texorpdfstring{Split-mask computational-basis readout}{Split-mask computational-basis readout}]\label{prop:split-mask-readout}
Let the logical computational-basis output bit be $y$, and let the final $X$ mask from the persistent matching-hidden split-QOTP frame, after applying the hidden common-node matching for the measured register, be split as
\begin{equation}
 x=x^{(1)}\oplus x^{(2)}\oplus x^{(R)}.
\end{equation}
Here $x^{(2)}$ denotes the second share aligned to the measured output by the common-node matching, and \(x^{(R)}\) denotes the hidden common-node refresh share in the recommended high-security variant.
If the measuring party obtains
\begin{equation}
 m=y\oplus x^{(1)}\oplus x^{(2)}\oplus x^{(R)},
\end{equation}
then the client recovers
\begin{equation}
 y=m\oplus x^{(1)}\oplus x^{(2)}\oplus x^{(R)}.
\end{equation}
Against any allowed coalition whose view does not determine the aligned final $X$ key, the output bit $y$ is information-theoretically hidden, provided the missing aligned key information is uniform and secret.
\end{proposition}

Only the $X$ component of the QOTP affects computational-basis readout. The $Z$ shares do not change computational-basis measurement outcomes, although they may still matter for state privacy before measurement or for measurements in other bases. Proposition \ref{prop:split-mask-readout} is the final-readout specialization of the persistent matching-hidden split-QOTP frame; state privacy comes from state privacy under persistent split-QOTP plus \(\epsilon_{\mathrm{key}}\) key hiding throughout the computation, while angle, structural and malicious-security claims require the separate assumptions stated below.

\begin{proposition}[Persistent matching-hidden split-QOTP state privacy under \(\epsilon_{\mathrm{key}}\) key hiding and shuffled $r$-share sign-randomized angle transcript ambiguity of Protocol~\ref{Protocol: 4}]\label{prop:protocol4-persistent-split-qotp-privacy}
Assume Protocol \ref{Protocol: 4} maintains a persistent split-QOTP and satisfies the \(\epsilon_{\mathrm{key}}\)-key-hiding matching compiler condition: for every allowed coalition $C\in\mathcal C_{\mathrm{P4}}$, the aligned Pauli key protecting every server-visible register is \(\epsilon_{\mathrm{key}}\)-close to uniform conditioned on $\mathsf{View}_C$ and the declared leakage. The claim has two separate components.
\begin{enumerate}
 \item \textbf{State privacy.} Then the coalition's quantum view is independent of the logical encrypted state up to the key-alignment error and the declared leakage. For the state-privacy component only, schematically,
\[
\Delta
\left(
\rho_{C,D}^{\Pi_4},
\operatorname{Sim}_C(L(D))
\right)
\leq
\epsilon_{\mathrm{key}}.
\]
 \item \textbf{Angle transcript ambiguity.} Separately, shuffled $r$-share sign-randomized angle sharing provides transcript-level angle ambiguity and transcript-level unlinkability under hidden signs, hidden grouping, finite-grid masking, auxiliary slots and common-node shuffling. This is a leakage-relative finite-grid transcript claim under hidden-sign, hidden-routing and hidden-grouping assumptions; it is not universal blindness and not a standalone full simulator-based quantum blindness theorem.
\end{enumerate}
The displayed trace-distance style bound applies to the state-privacy component only; no analogous bound is asserted for the angle transcript unless a separate transcript simulator and its error terms are defined.
The statement is conditioned on the declared leakage and the stated non-total-collusion coalition model. It is not a malicious-security, integrity or correctness-against-deviation theorem.
\end{proposition}
\begin{proof}[Proof sketch]
For any allowed coalition, the aligned Pauli key on each visible register is uniform or $\epsilon_{\mathrm{key}}$-close to uniform after conditioning on the coalition view and declared leakage. The Pauli twirl therefore makes the coalition's quantum marginal independent of the logical encrypted state up to that key-alignment error. The common-node permutations and refresh shares and auxiliary slots add matching and structural ambiguity for the angle-share transcript; this transcript-level statement is separate from the preceding trace-distance bound unless a separate transcript simulator and its error terms are explicitly defined. The proof does not address active deviations, which require a separate verification or authentication layer.
This is a conditional reduction statement rather than a composable security proof: if the stated key-hiding, hidden-routing and leakage-indistinguishability conditions hold, then the QOTP or transcript-ambiguity argument gives the claimed component guarantee.
\end{proof}

\paragraph{\texorpdfstring{Final split-mask readout primitive.}{Final split-mask readout primitive.}}
In the classical-client setting, output privacy is obtained at measurement time by reading the current persistent matching-hidden split-QOTP frame as in Proposition \ref{prop:split-mask-readout}: no allowed coalition may learn both the raw measurement outcome and the aligned full final $X$ key. The extra handoffs must be embedded in the ordinary routing pattern, and arbitrary measurement bases still require basis adaptation or an explicit delegated measurement subprotocol. 

\paragraph{Protocol \ref{Protocol: 4}: assumptions and guarantees.}
$ $

\textbf{Parties.} One classical client, two servers, and one common node; an optional extra measurement node may also be used.

\textbf{Assumptions.} The client generates all classical split-QOTP key-share tables, finite-grid angle shares, auxiliary slot schedules and Pauli-frame updates, and, in the recommended high-security variant, the common-node refresh shares used to satisfy the \(\epsilon_{\mathrm{key}}\)-key-hiding condition. Servers 1 and 2 are analyzed through the allowed coalitions of Assumption \ref{ass:protocol4-coalitions}. The common node honestly performs the prescribed rerouting, stores the matching permutations and hidden refresh shares, keeps them hidden from the computational servers and from coalitions not containing $R$, and is side-channel controlled. Coalitions containing $R$ may know the matching, but they do not also possess both computational value-share tables in the allowed-coalition model. In particular, the common node must not reveal routing correlations, relabeling information, refresh-share information or timing side information beyond the declared leakage. If output privacy is required, the final split-mask readout primitive above is used as the final measurement instance of the persistent matching-hidden split-QOTP frame.

\textbf{Adversarial model.} The privacy claims for this protocol are honest-but-curious, also called semi-honest: each server follows the prescribed quantum operations and routing handoffs, but may try to infer private angles, structure or outputs from its local transcript. The non-total-collusion and matching-hidden assumption is essential. Malicious deviations require a separate verification, authentication or composable-security layer introduced outside this protocol section.

\textbf{Leakage.} The servers learn $N$, the public parts of the circuit, the times at which handoffs occur, and whatever routing information is not hidden by the common node. They also learn their own local key-share tables, value-share transcripts and visible finite-grid angle shares. The declared leakage must specify any public operation classes, batch sizes, message sizes, timing, public routing template, grid precision, auxiliary-slot counts and visible handoff schedule. It must not include the common-node matching permutations unless that leakage is explicitly declared.

\textbf{Local views.} Server 1 sees the qubits it holds, its public instructions, its local Pauli-key shares, finite-grid value shares and visible operation slots and, in the computational-basis readout instantiation, the raw measurement string $m$. It does not see Server 2's private value-share table or the common-node matching. Server 2 sees the analogous local operations and its own shares, but not $m$ in the basic readout flow. The common node sees the physical routing pattern and relabeling information; the model assumes it does not reveal that information to either server beyond the prescribed labels.

\textbf{Correctness.} The finite-grid form of the shuffled $r$-share construction implies that the signed visible shares compose to the QOTP-corrected angle $\widetilde{\theta}_g$ for data slots and to a logically null or compiler-known value for auxiliary slots. When the final split-mask readout primitive is available, Proposition \ref{prop:split-mask-readout} shows that the client can recover the intended classical output from the measurement results.

\textbf{Privacy.} The main privacy statement of Protocol \ref{Protocol: 4} is Proposition \ref{prop:protocol4-persistent-split-qotp-privacy}: the state component is state privacy under persistent split-QOTP plus \(\epsilon_{\mathrm{key}}\) key hiding, while the angle component is transcript-level angle unlinkability from shuffled $r$-share sign-randomized finite-grid sharing. The trace-distance style part is the state-privacy statement; the angle statement is leakage-relative transcript-level unlinkability under the finite-grid hidden-sign, hidden-routing and hidden-grouping assumptions, not universal blindness. Protocol \ref{Protocol: 4} does not rely on client-side QOTP, since the classical client never locally applies or stores the encrypted quantum register. The routing-hidden sign-randomized construction of Proposition \ref{prop:sign-hidden-rz-blocks} is the $Z$-axis instance of a client-side or trusted-routing Pauli-axis angle-ambiguity primitive. In the classical-client two-computational-server plus common-node setting, the analogous effect is part of the base protocol assumptions: the servers and common node implement sign shares, auxiliary slots and routing masks without any allowed coalition learning both value shares and the full grouping information.

\textbf{Limitations.} Unlike Protocols \ref{Protocol: 1} and \ref{Protocol: 2}, this protocol does not provide client-side QOTP because the client never directly holds the encrypted register; instead, end-to-end state privacy is conditional on maintaining the persistent split-QOTP invariant together with the explicit \(\epsilon_{\mathrm{key}}\) key-hiding condition. Nor does Protocol \ref{Protocol: 4} inherit the optional non-Clifford gadget or QHE route discussed for Protocols \ref{Protocol: 2} and \ref{Protocol: 3}: the classical client cannot prepare the auxiliary quantum states or perform the client-side interaction required by those gadgets. It also relies on the non-total-collusion model, hidden matching, side-channel control, the explicit \(\epsilon_{\mathrm{key}}\) key-hiding condition, and, in the recommended high-security variant, the common-node Pauli refresh. If the full coalition $\{S_1,S_2,R\}$ forms, if matching information leaks, if structural leakage identifies the hidden grouping, or if the measuring party learns the aligned final key, the privacy guarantee degrades accordingly. Protocol \ref{Protocol: 4} still does not provide malicious security, integrity or correctness against active deviations; those require verification or authentication layers introduced later. With a larger number of servers, this setting could be adapted to federated quantum machine learning.

\begin{table*}[t]
\centering
\caption{\texorpdfstring{Operational overhead of the protocol hierarchy.}{Operational overhead of the protocol hierarchy.}}
\scriptsize
\setlength{\tabcolsep}{3pt}
\renewcommand{\arraystretch}{1.12}
\begin{tabular}{@{}L{0.13\textwidth}L{0.20\textwidth}L{0.19\textwidth}L{0.18\textwidth}L{0.13\textwidth}L{0.11\textwidth}@{}}
\toprule
Protocol & Extra quantum operations & Extra visible rotations & Routing/handoffs & Main overhead parameter & Main bottleneck \\
\midrule
Protocol \ref{Protocol: 1} & QOTP masks and local private gates & no extra visible rotations beyond target circuit & qubit transfers for local operations & number of client-server handoffs & client memory \(M\) \\
\midrule
Protocol \ref{Protocol: 2} & QOTP masks and local single-qubit gates & padding if structural hiding is used & routing permutations at handoffs & number of local layers & communication and port randomization \\
\midrule
Protocol \ref{Protocol: 3} & QOTP masks and sign-setting refresh & \(2r\) visible shares per protected Pauli-axis slot in masked-idle form & routing between share rounds & \(r,p,N_{\mathrm{slots}}\) & angle-sharing overhead \\
\midrule
Protocol \ref{Protocol: 4} with common-node refresh & split-QOTP plus common-node Pauli refresh & \(2r\) visible shares plus auxiliary slots where used & common-node matching plus refresh & \(r,p,m\) & trusted refresh capability and side-channel control \\
\midrule
Trap layer & verifier/dummy blocks & depends on verifier template & hidden intra-shot placement & verifier density \(L_v/L\) & verifier/data indistinguishability \\
\bottomrule
\end{tabular}
\label{tab:protocol-overhead}
\end{table*}

\subsection{\texorpdfstring{Fallback branches for restricted clients}{Fallback branches for restricted clients}}
After introducing the four protocol branches, the hierarchy can be compared with standard blind and verifiable delegation models. Depending on the client's quantum capabilities, different delegation branches provide different combinations of privacy, verifiability, and implementation cost. Table~\ref{tab:low-resource-branches} summarizes representative fallback branches for restricted clients.

\begingroup
\begin{table*}[t]
\centering
\caption{Representative fallback branches as a function of the client's available quantum resources. These branches are not additional numbered protocols in our framework; they indicate where the proposed QOTP/routing constructions sit relative to standard blind, verifiable, and measurement-based delegation models.}
\label{tab:low-resource-branches}
\begin{tabular}{p{0.25\textwidth} p{0.32\textwidth} p{0.35\textwidth}}
\toprule
\textbf{Client capability} & \textbf{Recommended branch} & \textbf{Security / privacy guarantee} \\
\midrule
Purely classical client, one server
&
Mahadev-style classical verification or computational QHE
&
Verification under computational assumptions such as LWE; no information-theoretic blindness by default. \\

Purely classical client, two servers that a priori should not communicate, plus a common node
&
Protocol \ref{Protocol: 4} or two-server UBQC / multi-prover blind computation
&
For Protocol \ref{Protocol: 4}, state privacy under persistent split-QOTP plus \(\epsilon_{\mathrm{key}}\) key hiding, and transcript-level angle unlinkability hold only under the stated non-total-collusion, hidden-matching, side-channel and leakage assumptions, with the common-node Pauli refresh in the recommended high-security variant; other branches have their own blindness or verifiability assumptions. \\

Client can prepare single-qubit states
&
UBQC-style branch
&
Blindness with limited quantum preparation under the assumptions of the chosen UBQC construction. \\

Client can only measure single qubits
&
Measurement-only BQC / VBQC branch
&
Blindness or verifiability when the server supplies suitable resource states. \\

Client can apply Pauli masks or handle a few qubits
&
Our QOTP/routing/sign-randomization branch, including Protocol \ref{Protocol: 3} where its assumptions apply
&
QOTP state privacy when fresh masks are client-controlled; angle or structure ambiguity under the declared leakage model. \\

Client has a single-qubit universal device
&
Protocol \ref{Protocol: 2} with $M=1$ for state privacy; Protocol \ref{Protocol: 3} only for the restricted angle-sharing branch when its routing assumptions are supplied
&
QOTP state privacy, with limited structural privacy unless additional padding or routing randomization is used. \\

Client has an $M$-qubit local register or $M$ independent single-qubit devices
&
Protocol \ref{Protocol: 2} for universal local single-qubit layers; Protocol \ref{Protocol: 3} for the restricted Pauli-mask/routing branch
&
State privacy plus structural privacy conditioned on the randomized compiler and leakage function. \\

Client has a full $M$-qubit device for the workload
&
Protocol \ref{Protocol: 1}
&
Strongest branch in our hierarchy, with direct encrypted delegation of Clifford layers and local handling of unsupported non-Clifford components. \\
\bottomrule
\end{tabular}
\end{table*}
\endgroup

\noindent\textbf{Position relative to standard blind and verifiable delegation.} The protocols in this work should not be read as replacements for UBQC, VBQC or Mahadev-style verification in their strongest security regimes. UBQC and VBQC provide stronger blindness or verifiability guarantees under their own resource assumptions, typically requiring specific quantum preparation, measurement capabilities or graph-state resources. Mahadev-style protocols give classical-client verification under computational assumptions. By contrast, our QOTP/routing hierarchy is optimized for a different axis: it makes explicit how much privacy can be retained when the client has partial quantum capability, restricted Pauli control, routing control, or only classical coordination under non-total-collusion. Thus the contribution is not a universal blind-computation theorem, but a resource-indexed protocol hierarchy with leakage-relative guarantees.

\begin{table*}[t]
\centering
\caption{\texorpdfstring{Compact comparison with standard blind, verifiable, homomorphic, and resource-indexed branches.}{Compact comparison with standard blind, verifiable, homomorphic, and resource-indexed branches.}}
\small
\setlength{\tabcolsep}{4pt}
\renewcommand{\arraystretch}{1.12}
\begin{tabular}{@{}L{0.22\textwidth}L{0.37\textwidth}L{0.33\textwidth}@{}}
\toprule
Branch & What it gives & What it does not claim \\
\midrule
UBQC/VBQC & stronger blindness/verifiability under their assumptions & not optimized for the same QOTP/routing hierarchy \\
\midrule
Mahadev-style & classical-client verification under computational assumptions & not information-theoretic blindness by default \\
\midrule
QHE/gadget routes & encrypted non-Clifford support under extra assumptions & higher overhead and not available to the Protocol \ref{Protocol: 4} classical client \\
\midrule
This hierarchy & explicit resource-indexed privacy/leakage tradeoff & not full blindness or malicious security by itself \\
\bottomrule
\end{tabular}
\label{tab:standard-comparison}
\end{table*}

Universal blind quantum computation is particularly relevant when the client can prepare single-qubit states from a small finite set and send them to the server. In that setting, the input, output, and computation pattern can be hidden from the server while requiring only limited quantum capability from the client~\cite{Universal_Blind}. Measurement-only blind computation covers a complementary regime: the client performs single-qubit measurements, while the server supplies suitable resource states~\cite{HayashiMorimae2015MeasurementOnly}. For verifiable blindness rather than detection-only checks, one should use a VBQC construction with its own proof, such as the Fitzsimons--Kashefi framework~\cite{FitzsimonsKashefi2017VBQC}.

For fully classical clients, the achievable guarantees change substantially. Mahadev-style protocols provide classical verification of quantum computation under computational assumptions such as LWE, but they should not be conflated with information-theoretic blindness unless additional mechanisms are introduced~\cite{Mahadev2018ClassicalVerification}. Computational QHE is another possible classical-client direction, but it changes the cryptographic assumptions and should be treated separately from the information-theoretic QOTP/routing branches considered here~\cite{Homomorphic_Circuits}.

Thus, the present framework should be read as a modular hierarchy indexed by client capability. The QOTP/routing/sign-randomization branches are useful in the low-resource quantum regime, but their privacy guarantees are always relative to the declared leakage function and to the assumed adversarial model.

\FloatBarrier
\section{\texorpdfstring{Trap-based detection layer}{Trap-based detection layer}}\label{sec: verification}
A separate issue is checking whether the server is actually applying the quantum operations it is instructed to perform. The server may deviate and return erroneous results even without learning the hidden quantum data.

Verification is difficult because, for large numbers of qubits, the system cannot simply be simulated classically, nor can properties of the problem always certify the solution. However, although there are other verification methods to consider \cite{Rational}, the protocols can include a trap-based detection check using verifier circuits and, when physically available, $\SWAP$ gates or equivalent hidden routing between data and verifier locations. This embedded statistical detection layer detects deviations under the stated trap-placement, non-adaptivity and indistinguishability assumptions; it is not used here as a complete malicious-security proof. All detection statements in this section are conditional component guarantees rather than composable malicious-security proofs: when the stated hidden-routing, placement-indistinguishability, leakage-indistinguishability and verifier-soundness assumptions hold, the verifier-density arguments below give the corresponding detection component guarantees.

We couple a checkable hidden embedded verifier instance to the data instance to be executed, while keeping their checked logical semantics separated inside the compiled execution. Let $T$ be the public observable template of the delegated computation. The template $T$ includes visible gate arities, layer structure, routing pattern, timing profile, communication schedule and measurement schedule. The hidden embedded formulation requires that the data block and every verifier block are independently randomized compilations of the same public observable template $T$, conditioned on the same declared leakage profile $L$.

The mixed intra-shot wire set is
\[
[N_{\mathrm{tot}}]=D\sqcup V_1\sqcup\cdots\sqcup V_k\sqcup A,
\]
where $D$ are data wires, $V_j$ are verifier blocks, and $A$ are auxiliary, dummy or padding wires. The client samples a private global placement/relabeling
\[
\pi:[N_{\mathrm{tot}}]\to [N_{\mathrm{tot}}],
\]
so the server observes only the permuted physical instance. Verification is embedded intra-shot: data wires, verifier wires, dummy wires and auxiliary wires are mixed inside the same compiled execution, rather than by alternating visibly separate data shots and verification shots. The server never observes which physical wires are data, verifier, dummy or auxiliary wires.

Each verifier is compiled as an independent global instance, with independence understood after conditioning on the declared leakage and on global placement consistency. The compiler randomness can be written as

\[
r =
(\pi_{\mathrm{glob}},
r_D,
r_{V_1},
\ldots,
r_{V_k},
r_A),
\]

where $\pi_{\mathrm{glob}}$ is the private global placement and $r_D,r_{V_1},\ldots,r_{V_k},r_A$ are the local randomizers for the data, verifier and auxiliary components. These local randomizers satisfy

\[
r_D\perp r_{V_1}\perp\cdots\perp r_{V_k}\perp r_A
\quad
\text{conditioned on } L
\text{ and on the global placement constraints.}
\]

Thus QOTP keys, Pauli frames, angle shares, sign masks, padding seeds, dummy choices, routing randomness and output masks are not reused across data and verifier instances, except for global consistency constraints required by the hidden placement and routing schedule.

\paragraph{Definition (hidden embedded verifier).}
Let $T$ be the public observable template of the delegated computation, including visible gate arities, layer structure, routing pattern, timing profile, communication schedule and measurement schedule. A hidden embedded verifier is a verifier instance $V_j$ compiled from $T$ on a private subset of logical wires, using fresh local randomization independent of the data instance and of all other verifier instances. Boundary interactions are completed with dummy, cancelling or auxiliary operations so that the server-visible transcript is identically distributed, or statistically close, to that of a data block conditioned on the declared leakage profile $L$.

The verifier compiler preserves the server-visible schedule, not necessarily the logical coupling pattern between data and verifier blocks. Boundary interactions of the restricted template are not omitted. If a template gate would connect a verifier wire $i\in S_j$ with a wire outside $S_j$, that public slot must be completed rather than dropped. Boundary interactions are completed by dummy, cancelling or auxiliary operations sampled from the same visible distribution as genuine interactions. More generally, completion may use dummy operations, auxiliary wires, cancelling sequences, compiled identities, or padding operations sampled from the same visible distribution as genuine interactions. Thus the public schedule, gate arities, timing, layer structure, routing profile and transcript distribution are preserved, while the logical data and verifier semantics remain separated and checkable.

\begin{proposition}[Verifier-location indistinguishability]\label{prop:verifier-location-indistinguishability}
Suppose that data and verifier blocks are compiled from the same observable template, with fresh independent local randomization, private placement, boundary completion and the same declared leakage profile. Then the server-visible distributions of data and verifier locations are identical, or $\varepsilon_{\mathrm{emb}}$-close, conditioned on the declared leakage profile. Formally,
\begin{equation}
\Delta
\bigl(
\mathsf{View}_{\mathrm{srv}}(D)\mid L,
\mathsf{View}_{\mathrm{srv}}(V_j)\mid L
\bigr)
\leq
\varepsilon_{\mathrm{emb}}.
\end{equation}
Consequently, any attack chosen as a function only of the server-visible view hits a verifier location with probability determined by the verifier density within the corresponding visible operation class.
\end{proposition}
The verifier subset may be described as the first $n_j$ logical wires only after private relabelling; the server never observes which physical locations correspond to data, verifier, dummy or auxiliary wires.

The corresponding protocol is executed on this single privately relabelled instance, so the visible transcript does not separate data, verifier, dummy and auxiliary roles. This makes it harder for the server to target only the data instance deliberately. The hiding reason is not mere visual similarity to the data circuit; it is the same observable template, fresh independent local randomization, private placement, boundary completion and the same declared leakage profile.

Finally, the client obtains the results of the data instance and the verifier. Because the verifier check is specified by one of the verifier modes below, a deviation affecting a verifier location is detected with some probability that depends on the verifier design and the attack model; some attacks can still go undetected if they act trivially on the verifier or preserve its output distribution. Thus, the embedded detection layer uses private relabelling, hidden routing or, when available, $\SWAP$ operations inside the compiled instance to mix logical roles without exposing the data/verifier partition.

The consistency check is the following.
\begin{enumerate}
 \item We prepare the $N$-qubit data block as one randomized compilation of $T$ conditioned on $L$.
 \item Add one or more verifier blocks of total $N'$ qubits, each checkable under one of the verifier modes below. Each verifier block is an independently randomized compilation of $T$ conditioned on the same declared leakage profile $L$. These verifier blocks will not be entangled with any of the $N$ qubits of the data block.
 \item Run the corresponding protocol for one combined intra-shot compiled instance on $N_{\mathrm{tot}}$ qubits, with the private placement $\pi$ mixing data, verifier, auxiliary, dummy and padding wires inside the same server-visible execution.
 \item Obtain the measurement results of the logical verifier blocks after undoing the private relabelling on the client side.
 \item If the logical verifier-block results do not match those of the selected verifier check, we conclude that the server deviated on at least one hidden verifier location. If they do match, we only conclude that the tested verifier locations were not caught deviating in that run.
\end{enumerate}

To state the detection probability more carefully, partition the server-visible attackable slots into visible operation classes $\tau\in\mathcal T$. A visible class includes the server-observable features contained in the declared leakage profile, such as gate label, arity, layer or timing, routing profile, measurement role, handoff or routing status, loss or routing metadata, and any other transcript metadata. Thus $\tau$ is not merely the logical gate type if the server can distinguish timing, routing or handoff status.
\begin{equation}
\begin{aligned}
L_v^{(\tau)}
&=
\#\{\text{verifier visible slots of class }\tau\},
\\
L_{\mathrm{tot}}^{(\tau)}
&=
\#\{\text{total visible slots of class }\tau\}.
\end{aligned}
\end{equation}
We also write $L_d^{(\tau)}=L_{\mathrm{tot}}^{(\tau)}-L_v^{(\tau)}$. The main verifier-hit probability is class-conditioned:
\begin{equation}
q_v^{(\tau)}
=
\frac{L_v^{(\tau)}}{L_{\mathrm{tot}}^{(\tau)}}.
\end{equation}
Since $0\leq L_v^{(\tau)}\leq L_{\mathrm{tot}}^{(\tau)}$, this is a probability in $[0,1]$ whenever $L_{\mathrm{tot}}^{(\tau)}>0$.
For a set $\mathcal T_{\mathrm{att}}$ of visible classes that the considered attack may target, the worst-case class density is
\begin{equation}
q_v^\star
=
\min_{\tau\in\mathcal T_{\mathrm{att}}:\,L_{\mathrm{tot}}^{(\tau)}>0}
q_v^{(\tau)}.
\end{equation}
If $L_v^{(\tau)}=0$ for an attacked visible class, then $q_v^{(\tau)}=0$ and attacks targeted exclusively at that class are not detected by this verifier layer, except through indirect effects or other checks.

The bound below holds in the ideal indistinguishability case of Proposition \ref{prop:verifier-location-indistinguishability}, within each visible operation class. With only approximate indistinguishability, one should add an $O(\varepsilon_{\mathrm{emb}})$ correction or state the corresponding statistical-distance loss explicitly. It also assumes the specified local attack model and, for repeated use, fresh independent randomization or an equivalent resampling of the adversary's effective target.

\begin{proposition}[Single-attack class-conditioned non-detection bound]\label{prop:verification}
Let $A$ be a local attackable visible slot of class $\tau$ chosen by the adversary, possibly as a function of the server's view. Suppose that, conditioned on the server's view and on the declared leakage profile, the client's randomization maps $A$ to a verifier location with probability $q_v^{(\tau)}$. Suppose further that any attack landing on a verifier location of class $\tau$ is detected by the classical verifier check with probability at least $\delta_\tau$. Then the idealized single-attack non-detection probability satisfies
\begin{equation}
 p_{\mathrm{nd}}^{(1)}(\tau)
 \leq
 1-\delta_\tau q_v^{(\tau)}.
\end{equation}
In the homogeneous uniform case in which all attackable locations are indistinguishable to the server and the randomization is uniform over $L_d$ data locations and $L_v$ verifier locations, one may only as a special case take
\begin{equation}
 q_v=\frac{L_v}{L_d+L_v},
\end{equation}
which gives
\begin{equation}
 p_{\mathrm{nd}}\leq 1-\delta\frac{L_v}{L_d+L_v}.
\end{equation}
The global symbol $q_v$ is therefore a shorthand for the homogeneous special case, while the main definition is $q_v^{(\tau)}$.
\end{proposition}

Proposition \ref{prop:verification} provides a detection bound under the stated randomization and indistinguishability assumptions. The values $q_v^{(\tau)}$ must be justified by the actual randomized placement and leakage model used in a concrete instantiation. 
The parameter $q_v^{(\tau)}$ is not a free parameter. It must be derived from the randomized placement procedure and from the server-view distribution within class $\tau$. If timing, routing, gate type, loss pattern, or interaction schedule distinguishes verifier locations from data locations, then the relevant class density may be much smaller, possibly zero.

\begin{proposition}[Multi-location non-detection bound]\label{prop:multi-location-verification}
Let $L$ be the total number of indistinguishable attackable locations in the combined data-plus-verifier circuit, let $L_v$ be the number of verifier or trap locations, and let $L_d=L-L_v$ be the number of data locations. Suppose the adversary attacks $t$ distinct locations in a single run before learning the hidden verifier/trap placement, or more generally suppose that, conditioned on the server's view, the attacked set is independent of the hidden placement. Suppose the $L_v$ verifier/trap locations are hidden uniformly among the $L$ attackable locations. If exactly $h$ of the attacked locations are verifier/trap locations, and each attacked verifier/trap location detects the deviation with probability at least $\delta$, with missed detections bounded by $(1-\delta)^h$, then the single-run non-detection probability satisfies
\begin{equation}
p_{\mathrm{nd}}(t)
\leq
\sum_{h=h_{\min}}^{h_{\max}}
\frac{
\binom{L_v}{h}
\binom{L-L_v}{t-h}
}{
\binom{L}{t}
}
(1-\delta)^h,
\end{equation}
where
\begin{equation}
h_{\min}=\max\{0,t-(L-L_v)\},
\qquad
h_{\max}=\min\{t,L_v\}.
\end{equation}
The hypergeometric factor is the probability that an attack on $t$ distinct locations hits exactly $h$ verifier/trap locations.

If $\delta=1$, any attacked verifier/trap location is detected, so non-detection occurs only if no verifier/trap location is hit. Therefore,
\begin{equation}
p_{\mathrm{nd}}(t)
\leq
\frac{
\binom{L-L_v}{t}
}{
\binom{L}{t}
}
\end{equation}
when $t\leq L-L_v$, and $p_{\mathrm{nd}}(t)=0$ when $t>L-L_v$. In this ideal-detection case,
\begin{equation}
\frac{
\binom{L-L_v}{t}
}{
\binom{L}{t}
}
=
\prod_{j=0}^{t-1}
\frac{L-L_v-j}{L-j}
\leq
\left(
1-\frac{L_v}{L}
\right)^t.
\end{equation}
\end{proposition}

The class-conditioned extension is obtained by applying the same hypergeometric argument separately inside each visible operation class. If the adversary attacks $t_\tau$ distinct slots of class $\tau$, and hidden placement is uniform or sufficiently symmetric within each class, then a product-form missed-detection bound is
\begin{equation}
p_{\mathrm{nd}}(\{t_\tau\})
\leq
\prod_{\tau}
\sum_{h=h_{\min}^{(\tau)}}^{h_{\max}^{(\tau)}}
\frac{
\binom{L_v^{(\tau)}}{h}
\binom{L_d^{(\tau)}}{t_\tau-h}
}{
\binom{L_{\mathrm{tot}}^{(\tau)}}{t_\tau}
}
(1-\delta_\tau)^h.
\end{equation}
Here
\begin{equation}
h_{\min}^{(\tau)}
=
\max\{0,t_\tau-L_d^{(\tau)}\},
\qquad
h_{\max}^{(\tau)}
=
\min\{t_\tau,L_v^{(\tau)}\}.
\end{equation}
This product form assumes that the classwise placement and missed-detection events admit the stated factorization; otherwise it must be replaced by the joint bound justified by the concrete verifier construction.

For $t=1$, the multi-location expression reduces to the same homogeneous-uniform intuition as Proposition \ref{prop:verification}, with $q_v=L_v/L$. Thus Proposition \ref{prop:verification} can be viewed as the one-location version, while Proposition \ref{prop:multi-location-verification} captures the probability of missing a larger attacked set in a single run.

The bound assumes that the adversary cannot identify verifier/trap locations before choosing the attacked set. The hidden placement is uniformly random, or at least sufficiently symmetric to induce the stated hypergeometric distribution, and the attacked locations are distinct. The attacked set must be chosen before the trap placement is revealed, or the trap placement must be independent of the adversary's choice conditioned on the server's view. The factor $(1-\delta)^h$ assumes that, conditional on hitting $h$ verifier/trap locations, the probability that all corresponding checks miss the deviation is at most $(1-\delta)^h$. If the verifier detections are correlated, this product form must be replaced by whatever joint missed-detection bound is justified by the verifier design. The result covers only attack models that the concrete verifier construction reduces to this effective location-attack model. 

If the multi-location experiment is repeated over $n$ independent runs, with fresh independent trap placements and an effective attack size $t_\ell$ in run $\ell$, then the product bound is
\begin{equation}
p_{\mathrm{nd}}^{(n)}
\leq
\prod_{\ell=1}^n
p_{\mathrm{nd}}^{(\ell)}(t_\ell),
\end{equation}
where each $p_{\mathrm{nd}}^{(\ell)}(t_\ell)$ is bounded by the hypergeometric expression above. If the same $t$ and the same parameters are used in every independent run, this becomes
\begin{equation}
p_{\mathrm{nd}}^{(n)}
\leq
p_{\mathrm{nd}}(t)^n.
\end{equation}
Without fresh independent randomization or an equivalent resampling of the adversary's effective target set, one should not exponentiate the single-run bound.

This bound is more informative than the single-location estimate when the adversary corrupts several locations in the same run. It shows that the non-detection probability decreases with the number of attacked locations, provided the verifier/trap locations remain indistinguishable from data locations. The multi-location estimate is therefore an improved detection estimate for the stated indistinguishable-location model. 

\noindent
For a sequence of $a$ independent local attacks under fresh randomization or an equivalent independence assumption, define
\begin{equation}
\delta^\star
=
\min_{\tau\in\mathcal T_{\mathrm{att}}}\delta_\tau.
\end{equation}
Then the simple class-worst-case estimate is
\begin{equation}
p_{\mathrm{nd}}
\lesssim
(1-\delta^\star q_v^\star)^a.
\end{equation}
This is a model-dependent bound for the effective local-attack model and independence assumptions above, not a universal malicious-security theorem. For attacks on several distinct locations in one run, Proposition \ref{prop:multi-location-verification} gives the corresponding single-run factor that should be used instead of exponentiating the one-location bound. Without the independence assumption, one should not simply exponentiate the single-attack bound.

\paragraph{Verifier modes.}
The embedded detection layer can use a hierarchy of verifier modes, depending on the circuit family and the client's resources. These modes keep the verifier hidden in the compiled execution and provide statistical detection under the placement and indistinguishability assumptions above; they are not a composable malicious-security proof.
\begin{enumerate}
 \item \textbf{Mode A: known-answer embedded verifiers.} This is the preferred mode when available. The client chooses inputs, private parameters, cancellations, masks or angles so that the decoded verifier outcome is known. For verifier block $V_j$, the check has the form
 
 \[
 \mathsf{Dec}_{r_j}(y_{V_j})=y_j^\star.
 \]
 
 Acceptance can be deterministic or nearly deterministic, depending on the noise model and on the chosen verifier construction.

 \item \textbf{Mode B: small classically simulable verifiers.} When known-answer instances are unavailable, the client may use small verifier blocks whose output distribution is classically computable. In that case the client knows
 
 \[
 p_{V_j}(y)=\Pr[C_{V_j}(x_j)=y].
 \]
 
 Verification then compares the observed verifier samples against the expected distribution rather than against a single deterministic answer.

 \item \textbf{Mode C: echo/symmetry verifiers.} For circuits with suitable layered, repetitive or symmetric structure, the verifier may use echo checks or cancelling companions, for example
 
 \[
 B^\dagger B=I
 \]
 
 or
 
 \[
 U_\ell^\dagger U_\ell=I.
 \]
 
 For QAOA, this should not be read as full translational symmetry of the problem Hamiltonian. Rather, QAOA has a layered and repetitive observable template,
 
 \[
 U_{\mathrm{QAOA}}(\gamma,\beta)
 =
 \prod_{\ell=1}^{p}
 e^{-i\beta_\ell H_M}
 e^{-i\gamma_\ell H_C}.
 \]
 
 Echo verifiers are suitable when a layer, sublayer or template block can be paired with an inverse or cancelling companion without changing the declared observable schedule.
\end{enumerate}

As a numerical example, suppose a single attacked verifier location is always detected, so $\delta=1$, and suppose we are in the homogeneous uniform case, so $q_v=L_v/(L_d+L_v)\approx N'/(N+N')$. If we also assume fresh independent randomization across $10^3$ runs, then Proposition \ref{prop:verification} gives a repeated non-detection bound of
\begin{equation}
 (1-q_v)^{10^3}.
\end{equation}
If $L_v=L_d$, then $q_v=1/2$ and the bound is $2^{-1000}$. If $L_v=L_d/10$, then $q_v=1/11$ and the bound is $(10/11)^{1000}\approx 4\times 10^{-42}$. Even with $L_v=L_d/100$, we have $q_v=1/101$ and the bound is $(100/101)^{1000}\approx 4.8\times 10^{-5}$. These numbers are illustrative only and rely on the independence assumptions above. 

For a short multi-location illustration, in the ideal case $\delta=1$, if $L_v=L/10$ and an adversary attacks $t=5$ locations in a single run, then
\begin{equation}
p_{\mathrm{nd}}(5)
\leq
\left(\frac{9}{10}\right)^5\approx 0.6
\end{equation}
by the simple product upper bound, with the exact hypergeometric expression slightly smaller than $(9/10)^5$. This example is only illustrative and relies on uniform indistinguishable placement.

In addition, this consistency layer can be interleaved with the computation, so the trap-qubit check can be evaluated alongside the delegated run. 

Hidden embedded subcircuit verification can increase detection of malicious deviations by making verifier and data locations indistinguishable within the declared leakage profile, under the non-adaptive or placement-independent attack assumptions above. Detection probability depends on the density of verifier slots in each visible operation class and on the soundness of the chosen verifier mode.

This mechanism should therefore be understood as an embedded statistical detection layer, not as a complete malicious-security proof. Full malicious security would require a trap/authentication framework with its own soundness proof, or a simulator-based verifiable-delegation construction.

\paragraph{\texorpdfstring{Authenticated QOTP as a high-resource variant.}{Authenticated QOTP as a high-resource variant.}}
The trap-based layer gives statistical detection under the assumptions above, but it is not quantum authentication. A sufficiently capable client could instead replace bare QOTP blocks with authenticated quantum encryption, gaining an integrity check at the cost of extra qubits, encoding and decoding operations, and a different security analysis; Appendix \ref{app:authenticated-qotp} records the high-resource variant.

\section{Examples with quantum algorithms}\label{sec: examples}
This section applies the earlier protocols and primitives to three algorithmic templates: Grover's algorithm, QAOA and a QNN. For simplicity, we do not include the trap-based detection layer here. The examples should be read as delegation schedules together with the assumptions stated in the previous sections, rather than as stand-alone security proofs.

\subsection{Grover's algorithm}
Grover's algorithm \cite{Grover} is a quantum search algorithm based on repeatedly applying two operators so that the probability of measuring the target state or states is maximized. The two repeated operators are Grover's diffusion operator $U_s$, which applies an inversion with respect to the mean, and the search oracle $U_\omega$, which marks the target state or states with a negative sign. Fig. \ref{fig: Grover circuit} shows Grover's circuit for the states $\ket{101}$ and $\ket{110}$.

\begin{figure}[ht]
 \centering
 \begin{tikzpicture}
 \node[scale=0.9]{
 \begin{quantikz}
 \ket{0} & \gate{H} & \ctrl{2}\gategroup[3,steps=2,style={inner sep=2pt}]{$U_\omega$}& \control{} & \gate{H}\gategroup[3,steps=5,style={inner sep=2pt}]{$U_s$} & \gate{X} & \ctrl{2} & \gate{X} & \gate{H} &\qw \cdots & \meter{}\\
 \ket{0} & \gate{H} & \ocontrol{}& \control{} & \gate{H} & \gate{X} & \control{} & \gate{X} & \gate{H} & \qw\cdots & \meter{}\\
 \ket{0} & \gate{H} & \control{}& \octrl{-2} & \gate{H} & \gate{X} & \gate{Z} & \gate{X} & \gate{H} & \qw\cdots & \meter{}
 \end{quantikz}
 };
 \end{tikzpicture}
 \caption{\texorpdfstring{Grover circuit searching for the states $\ket{101}$ and $\ket{110}$.}{Grover circuit searching for the states 101 and 110.}}
 \label{fig: Grover circuit}
\end{figure}

\begin{figure}[ht]
 \centering
 \begin{tikzpicture}
 \node[scale=0.75]{
 \begin{quantikz}
 \ket{0} & \gate{H}\gategroup[2,steps=2,style={inner sep=2pt, fill=blue!20},background]{} & \qw\slice{$K$}& \ctrl{2}\gategroup[3,steps=1,style={inner sep=2pt, fill=red!20},background]{}\slice{$K^{-1}$}&\qw&\qw\slice{$K$}& \ctrl{1}\gategroup[2,steps=1,style={inner sep=2pt, fill=red!20},background]{}\slice{}&\qw\gategroup[3,steps=1,style={inner sep=2pt, fill=blue!20},background]{}\slice{}&\qw\slice{}&\swap{1}\slice{}\gategroup[2,steps=1,style={inner sep=2pt, fill=blue!20},background]{}&\ctrl{2}\gategroup[3,steps=3,style={inner sep=2pt, fill=red!20},background]{}&\ctrl{1}&\ctrl{2}\slice{$K^{-1}$}&\gate{X}\gategroup[2,steps=1,style={inner sep=2pt, fill=blue!20},background]{}\slice{$K$}&\qw \\
 \ket{0} & \gate{H} & \gate{X} &\control{}& \gate{X}\gategroup[2,steps=2,style={inner sep=2pt, fill=blue!20},background]{} &\swap{1}&\control{} &\qw &\ctrl{1}\gategroup[2,steps=1,style={inner sep=2pt, fill=red!20},background]{}&\targX{}&\qw&\control{}&\control{}&\qw&\qw\\
 \ket{0} & \gate{H}\gategroup[1,steps=2,style={inner sep=2pt, fill=blue!20},background]{} & \qw&\control{}&\gate{X} &\targX{}&\qw&\qw&\control{}&\qw&\control{}&\qw&\control{}&\qw&\qw
 \end{quantikz}
 };
 \end{tikzpicture}
 \centering
 \begin{tikzpicture}
 \node[scale=0.7]{
 \begin{quantikz}
 & \gate{H}\gategroup[3,steps=5,style={inner sep=2pt, fill=red!20},background]{}& \gate{X} & \ctrl{2} & \gate{X} &\gate{H} & \qw\cdots\slice{$K^{-1}$} & \meter{}\\
 & \gate{H} & \gate{X} & \control{} & \gate{X} & \gate{H} &\qw \cdots & \meter{}\\
 & \gate{H} & \gate{X} & \control{} & \gate{X} & \gate{H} &\qw \cdots & \meter{}
 \end{quantikz}
 };
 \end{tikzpicture}
 \caption{\texorpdfstring{Delegated Grover circuit searching for the states $\ket{101}$ and $\ket{110}$. Blue regions are executed by the client and red regions by the server. The dividers indicate qubit handoffs and routing updates and whether the register is encrypted ($K$) or decrypted ($K^{-1}$).}{Delegated Grover circuit searching for the states 101 and 110. Blue regions are executed by the client and red regions by the server. The dividers indicate qubit handoffs and routing updates and whether the register is encrypted or decrypted.}}
 \label{fig: Grover applied circuit}
\end{figure}

In this example, the protected information is the oracle $U_\omega$ that marks the desired state. Because the oracle is mainly structural and the problem information is not encoded in continuous angles, this structure must be protected. The schedule below is therefore schematic. In the Grover schedule, any occurrence of a private $\CCZ$ should be read either as a local three-qubit operation by a sufficiently capable Protocol \ref{Protocol: 1} client with $M\geq 3$, or as a compiled Clifford-plus-local-single-qubit subroutine for a Protocol \ref{Protocol: 2} client with universal single-qubit devices. It should not be read as a direct server-side non-Clifford evaluation on QOTP-encrypted data. For Protocol \ref{Protocol: 2}, the server receives only the public Clifford layers of the chosen $\CCZ$ decomposition, while the client receives the relevant qubits for the local single-qubit gates, applies them, freshly encrypts, and returns the qubits. This increases gate count, communication rounds and re-encryption steps. One two-qubit handoff schedule for this compiled-client interpretation is as follows, represented in Fig. \ref{fig: Grover applied circuit}:
\begin{enumerate}
 \item The client creates the first two qubits and applies $H$ gates to each one. The client also applies the $X$ gate to the second qubit.
 \item The client encrypts the qubits and sends them.
 \item The client creates the third qubit and applies an $H$ gate to it.
 \item The client encrypts the qubit and sends it.
 \item The schedule reaches an \ApplyPrivateCCZ{} block, interpreted according to the local-or-compiled meaning above.
 \item The client receives back the last two qubits, decrypts them, applies an $X$ gate to each one, encrypts them and updates the routing permutation for the two returned logical wires before returning them to the server.
 \item The server applies $\CZ$ between the first two qubits and sends them to the client.
 \item The client does nothing and returns the qubits.
 \item The server applies $\CZ$ between the last two qubits and sends them to the client.
 \item The client receives back the first two qubits and updates the logical-to-physical port assignment before returning them to the server.
 \item The server applies two $\CZ$ gates (which eliminate the two recent operations), the schedule reaches another \ApplyPrivateCCZ{} block under the same local-or-compiled interpretation, and sends the first two qubits to the client.
 \item The client decrypts the qubits, applies an $X$ gate to the first one, encrypts them and sends them to the server.
 \item The server applies the public Clifford parts of the Grover diffusion operator, with each non-Clifford block again realized through the local-or-compiled \ApplyPrivateCCZ{} interpretation.
 \item The process is repeated until the final measurement.
\end{enumerate}

\paragraph{Delegation map for Grover.}
The private component is the oracle structure $U_\omega$ together with the placement of the padding and any physically available $\SWAP$ gates or rerouting operations that hide it. In the schedule above, the client performs the local $H$, $X$, encryption, decryption, local non-Clifford or local single-qubit decomposition steps, and rerouting steps. The server performs only the public Clifford blocks delegated to it. The QOTP key-update rules of Proposition \ref{prop:clifford-update} are sufficient for the delegated $\CZ$ and other Clifford gates because they are Clifford. They are \emph{not} sufficient for $\CCZ$, which is non-Clifford by Assumption \ref{ass:clifford-only}. Correctness is obtained for the locally executed or compiled circuit, but structural privacy of the oracle is not automatic. If the location, controls or pattern of these $\CCZ$ blocks is private, the delegated Clifford skeleton of the decomposition must be hidden by the randomized structural-privacy compiler, padding, routing permutations or another blind/private subroutine. The compilation solves the encrypted-evaluation obstruction but does not by itself solve structural leakage. If no such structural hiding is supplied, the Grover example remains a delegation template rather than a complete structural-privacy proof.

\subsection{QAOA}
In QAOA, the circuit also repeats two operators for a chosen depth, with parameters optimized to increase the probability of low-cost bit strings. The first operator is a cost Hamiltonian operator $U_H$, which for a QUBO problem applies $R_Z$ and $R_{ZZ}$ gates. The second operator is the mixing operator $U_X$, which is composed of several $R_X$ gates. Thus, the circuit has the form shown in Fig. \ref{fig: QAOA circuit}.
\begin{figure}[ht]
 \centering
 \begin{tikzpicture}
 \node[scale=0.8]{
 \begin{quantikz}[transparent]
 \ket{0} & \gate{H} & \gate{R_Z(\theta_1)}\gategroup[3,steps=4,style={inner sep=2pt}]{$U_H$}& \gate[2]{R_{ZZ}(\theta_4)} & \gate[3,label style={yshift=0.3cm}]{R_{ZZ}(\theta_5)} & \qw& \gate{R_X(\phi_1)}\gategroup[3,steps=1,style={inner sep=2pt}]{$U_X$} &\qw \cdots & \meter{}\\
 \ket{0} & \gate{H} & \gate{R_Z(\theta_2)} & & \linethrough & \gate[2]{R_{ZZ}(\theta_6)} & \gate{R_X(\phi_2)} & \qw\cdots & \meter{}\\
 \ket{0} & \gate{H} & \gate{R_Z(\theta_3)}& \qw & & & \gate{R_X(\phi_3)} & \qw\cdots & \meter{}
 \end{quantikz}
 };
 \end{tikzpicture}
 \caption{\texorpdfstring{Three-qubit QAOA circuit.}{Three-qubit QAOA circuit.}}
 \label{fig: QAOA circuit}
\end{figure}

Here, the circuit structure is public only if the QAOA instance itself is embedded in a public graph or template. If the graph, topology or sparsity pattern of the QUBO is private, then one must hide it with a public superset template, dummy edges, masked idle rotations rather than directly recognizable zero-angle dummies, padding, or another equivalent mechanism. Under that caveat, it is sufficient to protect the $\theta$ and $\phi$ angles when they are private. For this, we use the decomposition of Fig. \ref{fig:rzz-decomposition}, where the private parameter remains in a single-qubit $R_Z$ gate and the entangling $\CNOT$ gates can be delegated publicly. The two-qubit client schedule is as follows, represented in Figs. \ref{fig: QAOA applied circuit} (original) and \ref{fig: QAOA optimized applied circuit} (optimized):

\begin{enumerate}
 \item The client creates the first two qubits and applies the corresponding $H$ and $R_Z$ gates to them, then encrypts and sends them to the server.
 \item The client creates the third qubit and applies the corresponding $H$ and $R_Z$ gates to it while the server in parallel applies the $\CNOT$ gate to the first two qubits. The client encrypts its qubit, sends it, and requests qubit $2$ from the server.
 \item The client decrypts the second qubit and applies the $R_Z$ gate, while the server applies the $\CNOT$ gate. Then, the client encrypts and sends the qubit.
 \item The server applies the $\CNOT$ gate.
 \item The client requests the first qubit, decrypts it and applies the $R_X$ gate, while the server applies the $\CNOT$ gate. The client encrypts its qubit, sends it, and then requests the third qubit.
 \item The client decrypts the third qubit and applies the $R_Z$ gate. Then, the client encrypts and sends the qubit.
 \item The server applies the $\CNOT$ gate and sends the two last qubits to the client.
 \item The client decrypts the qubits and applies the $R_X$ gates.
 \item The process is repeated for each layer, until the final measurement.
\end{enumerate}

\begin{figure}[ht]
 \centering
 \begin{tikzpicture}
 \node[scale=0.8]{
 \begin{quantikz}[transparent]
 \ket{0} & \gate{H}\gategroup[2,steps=2,style={inner sep=2pt, fill=blue!20},background]{} & \gate{R_Z(\theta_1)}\slice{$K$}& \ctrl{1}\gategroup[2,steps=1,style={inner sep=2pt, fill=red!20},background]{}\slice{$K^{-1}$} &\qw \slice{$K$} &\ctrl{1}\gategroup[3,steps=2,style={inner sep=2pt, fill=red!20},background]{}&\ctrl{2}\slice{$K$}&\qw&\ctrl{2}\gategroup[3,steps=2,style={inner sep=2pt, fill=red!20},background]{}&\qw\slice{$K^{-1}$}&\qw\slice{$K$}&\qw\slice{$K^{-1}$}& \gate{R_X(\phi_1)}\gategroup[3,steps=1,style={inner sep=2pt, fill=blue!20},background]{} & \qw\cdots\slice{} & \meter{}\\
 \ket{0} & \gate{H} & \gate{R_Z(\theta_2)} & \targ{} & \gate{R_Z(\theta_4)}\gategroup[1,steps=1,style={inner sep=2pt, fill=blue!20},background]{} &\targ{}&\qw&\qw&\qw&\ctrl{1}&\qw&\ctrl{1}\gategroup[2,steps=1,style={inner sep=2pt, fill=red!20},background]{} & \gate{R_X(\phi_2)} & \qw\cdots & \meter{}\\
 \ket{0} & \gate{H}\gategroup[1,steps=2,style={inner sep=2pt, fill=blue!20},background]{} & \gate{R_Z(\theta_3)}& \qw&\qw& \qw& \targ{}&\gate{R_Z(\theta_5)}\gategroup[1,steps=1,style={inner sep=2pt, fill=blue!20},background]{}& \targ{}&\targ{}&\gate{R_Z(\theta_6)}\gategroup[1,steps=1,style={inner sep=2pt, fill=blue!20},background]{}&\targ{} & \gate{R_X(\phi_3)} &\qw \cdots & \meter{}
 \end{quantikz}
 };
 \end{tikzpicture}
 \caption{\texorpdfstring{Three-qubit delegated QAOA circuit. Blue regions are executed by the client and red regions by the server.}{Three-qubit delegated QAOA circuit. Blue regions are executed by the client and red regions by the server.}}
 \label{fig: QAOA applied circuit}
\end{figure}

\paragraph{Delegation map for QAOA.}

The private gates are the parameterized rotations $R_Z(\theta_i)$, and also the $R_X(\phi_i)$ gates if the mixer angles are regarded as confidential. If the client can apply $R_Z(\theta)$ locally, that remains the simplest option: the client performs the private single-qubit rotation, while the server performs only public $\CNOT$ gates on encrypted data. This is exactly the Clifford setting of Proposition \ref{prop:clifford-update}. If the client cannot apply the private $R_Z$ rotation but can impose hidden $X$ keys and routing permutations, the private $R_Z$ layer can instead be implemented using the $Z$-axis instance of the finite-grid routing-hidden sign-randomized Pauli-axis rotation construction of Proposition \ref{prop:sign-hidden-rz-blocks}. The private parameters are first compiled to labels in $\mathbb{Z}_q$ under the public precision model. Under the convention fixed in Sec. \ref{sec: techniques},

\begin{equation}
 \CNOT_{12}(I\otimes R_Z(\theta))\CNOT_{12}=R_{ZZ}(\theta),
\end{equation}

so the private $R_{ZZ}$ parameter is implemented through an $R_Z(\theta)$ sandwiched by delegated public $\CNOT$ gates. The $R_Z(\theta)$ part is either applied locally by the client or implemented by the $Z$-axis instance of the finite-grid sign-randomized sharing construction under its leakage assumptions. If only a subset of qubits receives private rotations in a QAOA layer, idle real circuit qubits may be used as dummy positions with masked identity rotations $R_Z(-\eta)R_Z(\eta)$, avoiding additional qubit overhead. This increases the number of visible $R_Z$ calls, handoffs, re-encryptions and routing operations. If the graph structure or the placement or timing of the $R_Z(\theta)$ rotations is private, this schedule must also be combined with a hiding mechanism for the edge pattern, not only for the rotation angles.

\noindent\textbf{Example leakage function for padded QAOA/QUBO.} For a private QUBO instance with graph \(G=(V,E)\), coefficients \(J_{uv}\), local fields \(h_v\), QAOA depth \(P\), and private angles \((\gamma_\ell,\beta_\ell)\), a possible declared leakage function is
\[
L_{\mathrm{QAOA}}(D)=
\bigl(
|V|,\,
G_{\mathrm{pub}},\,
P,\,
p,\,
\text{public layer calendar},\,
\text{visible operation classes},\,
\text{message sizes},\,
\text{number of shots/iterations}
\bigr).
\]
If the optimizer transcript is not padded, \(L_{\mathrm{QAOA}}\) must also include the visible optimizer transcript, including iteration count, stopping time, batch identities and any parameter-reuse pattern visible to the server. The leakage function does not include \(E\), private coefficients \(J_{uv},h_v\), private compiled angle labels, masks \(\eta\), real/dummy edge indicators, QOTP keys, split-key alignments, share groupings, routing permutations, or output masks unless deliberately declared as leakage.

\begin{figure}[ht]
 \centering
 \begin{tikzpicture}
 \node[scale=0.8]{
 \begin{quantikz}[transparent]
 \ket{0} & \gate{H;\,R_Z(\theta_1)}\gategroup[2,steps=1,style={inner sep=2pt, fill=blue!20},background]{} & \ctrl{1}\gategroup[2,steps=1,style={inner sep=2pt, fill=red!20},background]{} &\ctrl{2}\gategroup[3,steps=1,style={inner sep=2pt, fill=red!20},background]{}&\ctrl{1}\gategroup[2,steps=1,style={inner sep=2pt, fill=red!20},background]{}&\ctrl{2}\gategroup[3,steps=2,style={inner sep=2pt, fill=red!20},background]{}& \gate{R_X(\phi_1)}\gategroup[1,steps=1,style={inner sep=2pt, fill=blue!20},background]{} &\qw&\qw&\qw&\qw \cdots & \meter{}\\
 \ket{0} & \gate{H;\,R_Z(\theta_2)} & \targ{} &\gate{R_Z(\theta_4)}\gategroup[1,steps=1,style={inner sep=2pt, fill=blue!20},background]{} &\targ{}&\qw&\ctrl{1}&\qw&\ctrl{1}\gategroup[2,steps=1,style={inner sep=2pt, fill=red!20},background]{} & \gate{R_X(\phi_2)}\gategroup[2,steps=1,style={inner sep=2pt, fill=blue!20},background]{} & \qw\cdots & \meter{}\\
 \ket{0} &\qw &\gate{H;\,R_Z(\theta_3)}\gategroup[1,steps=1,style={inner sep=2pt, fill=blue!20},background]{} &\targ{} &\gate{R_Z(\theta_5)}\gategroup[1,steps=1,style={inner sep=2pt, fill=blue!20},background]{}& \targ{}&\targ{}&\gate{R_Z(\theta_6)}\gategroup[1,steps=1,style={inner sep=2pt, fill=blue!20},background]{}&\targ{} & \gate{R_X(\phi_3)} &\qw \cdots & \meter{}
 \end{quantikz}
 };
 \end{tikzpicture}
\caption{\texorpdfstring{Optimized three-qubit delegated QAOA circuit using parallelized operations. Blue regions are executed by the client and red regions by the server.}{Optimized three-qubit delegated QAOA circuit using parallelized operations. Blue regions are executed by the client and red regions by the server.}}
 \label{fig: QAOA optimized applied circuit}
\end{figure}

In Fig. \ref{fig: QAOA optimized applied circuit}, each combined blue label $H;\,R_Z(\theta_i)$ is read from left to right: the client first applies $H$ and then $R_Z(\theta_i)$.

\subsection{QNN}
For a standard quantum neural network, one may want to protect the input data for each training step. These data are the initial encoding angles of each qubit, so the rest of the circuit can be executed by the server when the trainable weights are public. However, if the weights of the model must also be hidden, this can be done by having the client apply the corresponding single-qubit gates. The two-qubit client schedule is as follows, represented in Fig. \ref{fig: QNN circuit}:
\begin{enumerate}
 \item The client generates two qubits and applies $R_X$ and $R_Y$ gates to them to encode the data and the first layer of weights. It encrypts them and sends them to the server.
 \item The client generates a third qubit and applies the corresponding $R_X$ and $R_Y$ gates to it, while the server applies the $\CNOT$ gate. The client encrypts and sends its qubit, while requesting the first one.
 \item The client decrypts the first qubit and applies the $R_Y$ gate, while the server applies the $\CNOT$ gate. The client freshly encrypts and sends its qubit, while requesting the last two qubits.
 \item The client decrypts the last two qubits and applies the $R_Y$ gates. The client freshly encrypts and sends those qubits.
 \item The server applies the last two $\CNOT$ gates.
 \item The layer pattern is repeated until the final measurement.
\end{enumerate}

\paragraph{Delegation map for the QNN example.}

The private data are the input-encoding angles $x_i$, and the weights $\omega_i$ are also private whenever the model parameters must be hidden. In the displayed schedule, local client rotations remain the cleanest option: the client performs the parameterized $R_X$ and $R_Y$ gates as local single-qubit operations, while the server performs only public entangling $\CNOT$ gates on encrypted data. Hence the delegated part again lies inside the Clifford regime covered by Proposition \ref{prop:clifford-update}. For restricted clients, private one-qubit Pauli-axis rotation layers, namely $R_X$, $R_Y$ and $R_Z$, can use the same finite-grid sign-randomized primitive by choosing the hidden Pauli mask that anticommutes with the corresponding rotation axis. The resulting angle ambiguity remains conditional on the same hidden-sign, hidden-grouping, finite-grid and routing assumptions as in the basic primitive. No generic direct non-Clifford encrypted evaluation by the server is assumed; the overhead appears as additional visible rotation calls, handoffs, re-encryptions and routing operations whenever a private parameterized rotation is delegated this way.

\paragraph{Repeated-shot and optimizer leakage for variational workloads.}
QAOA, QNN and other variational workloads require an additional transcript-leakage check. Even if every shot uses fresh QOTP keys and fresh angle randomization, the sequence of requested parameters may leak information through optimizer correlations, repeated parameter values, stopping rules, batch identities, iteration counts, cost-function structure or ansatz structure. Therefore the declared leakage function for a variational workload must include the optimization transcript, or the implementation must add a padding and randomization layer that makes transcripts distributionally indistinguishable inside the selected workload class.

\begin{table*}[t]
\centering
\caption{\texorpdfstring{Repeated-shot and optimizer-transcript leakage in variational workloads.}{Repeated-shot and optimizer-transcript leakage in variational workloads.}}
\scriptsize
\setlength{\tabcolsep}{3pt}
\renewcommand{\arraystretch}{1.12}
\begin{tabular}{@{}L{0.18\textwidth}L{0.27\textwidth}L{0.25\textwidth}L{0.22\textwidth}@{}}
\toprule
Leakage source & Why it matters & Mitigation & If not mitigated \\
\midrule
Repeated parameter values & Allows cross-shot alignment of candidate angle sets & Fresh masks, fresh routing, randomized ordering & Include parameter repetition pattern in \(L(D)\) \\
\midrule
Optimizer trajectory & Parameter updates may reveal cost landscape or private instance structure & Batch padding, delayed reporting, transcript smoothing & Include optimizer transcript in \(L(D)\) \\
\midrule
Stopping rule & Early stopping may leak convergence properties of private instance & Fixed iteration budget or padded stopping & Include stopping time in \(L(D)\) \\
\midrule
Batch identities & Lets server align the same logical edge/qubit across iterations & Randomized batch labels and dummy batches & Include batch identity leakage \\
\midrule
Timing/loss patterns & May reveal real vs dummy operations or hidden matching & Constant-time routing, padded message sizes & Include timing/loss metadata \\
\midrule
Port labels/routing latency & Can reveal hidden grouping or \(\sigma_R\) & Port relabeling, synchronized handoffs & Include matching leakage \\
\midrule
Visible zero/dummy structure & May reveal absent edges or idle qubits & Masked-idle rotations \(R_A(-\eta)R_A(\eta)\) & Include dummy/real distinction \\
\bottomrule
\end{tabular}
\label{tab:variational-transcript-leakage}
\end{table*}

\begin{figure}[ht]
 \centering
 \begin{tikzpicture}
 \node[scale=0.8]{
 \begin{quantikz}
 \ket{0} & \gate{R_X(x_1)}\gategroup[2,steps=2,style={inner sep=2pt, fill=blue!20}, background]{}& \gate{R_Y(\omega_1)} &\ctrl{1}\gategroup[2,steps=1,style={inner sep=2pt, fill=red!20}, background]{}&\qw&\gate{R_Y(\omega_4)}\gategroup[1,steps=1,style={inner sep=2pt, fill=blue!20}, background]{}&\qw& \ctrl{1}\gategroup[3,steps=2,style={inner sep=2pt, fill=red!20}, background]{}&\qw&\qw\cdots & \meter{}\\
 \ket{0} & \gate{R_X(x_2)} & \gate{R_Y(\omega_2)} & \targ{} &\qw& \ctrl{1}\gategroup[2,steps=1,style={inner sep=2pt, fill=red!20}, background]{} & \gate{R_Y(\omega_5)}\gategroup[2,steps=1,style={inner sep=2pt, fill=blue!20}, background]{}&\targ{} & \ctrl{1}&\qw \cdots & \meter{}\\
 \ket{0} & \qw&\qw& \gate{R_X(x_3)}\gategroup[1,steps=2,style={inner sep=2pt, fill=blue!20}, background]{}& \gate{R_Y(\omega_3)} & \targ{} &\gate{R_Y(\omega_6)} & \qw&\targ{} & \qw\cdots & \meter{}
 \end{quantikz}
 };
 \end{tikzpicture}
 \caption{\texorpdfstring{Three-qubit delegated QNN circuit. Blue regions are executed by the client and red regions by the server.}{Three-qubit delegated QNN circuit. Blue regions are executed by the client and red regions by the server.}}
 \label{fig: QNN circuit}
\end{figure}

\section{Security assumptions and limitations}\label{sec: limitations}
The protocols above are intentionally modular, and their privacy claims should be read together with the following assumptions and limitations.
\begin{itemize}
 \item The correctness discussion uses ideal channels and noiseless storage/transmission. Physical noise, device calibration errors and fault-tolerant implementation issues are outside the model.
 \item Unless explicitly strengthened, privacy is honest-but-curious and leakage-dependent: in the strict model the server follows the prescribed channels and retains its allowed transcript, while in the purified model it may keep compatible ancillas. Malicious deviations require a separate proof or mechanism and are treated here only through the trap-based detection layer and any optional authentication mechanism supplied separately.
 \item The QOTP gives perfect secrecy of a quantum state only when the Pauli keys are uniformly random, kept secret from the server, and not reused in a way that correlates different encrypted states.
 \item Simple Pauli-frame updates apply to delegated Clifford gates such as $H$, $S$, $\CNOT$ and $\CZ$. In the base protocols, the server is not asked to evaluate non-Clifford gates directly on QOTP-encrypted data; such gates are handled by local client execution, by a Clifford-plus-local-single-qubit decomposition, or by an explicitly supplied additional primitive.
 \item Optional gadget-based or QHE-style non-Clifford extensions are relevant only to Protocols \ref{Protocol: 2} and \ref{Protocol: 3} when the client is given the additional quantum ability to prepare auxiliary states or participate in the required interaction. They are not part of Protocol \ref{Protocol: 4}, whose client is classical and below that capability threshold.
 \item Structural privacy from padding gates, redundant gates, physical $\SWAP$ gates, routing permutations and port randomization is always relative to the declared leakage model and to the randomized compiler that produces the delegated circuit. Circuit size, number of qubits, public gate locations, timing information and communication patterns may still leak information unless they are also padded.
 \item The primitive is limited to one-qubit Pauli-axis rotations, $R_X$, $R_Y$ and $R_Z$, and does not by itself provide blindness for arbitrary non-Clifford gates or arbitrary rotations about non-Pauli axes. It provides finite-grid, leakage-dependent angle ambiguity only under its stated leakage assumptions: fresh independent hidden signs, hidden effective Pauli keys, hidden grouping through routing permutations or port randomization, finite-grid masks, distributionally masked idle rotations and fresh per-shot randomness. The precision parameter $p$ is an implementation and leakage parameter, while $r$ is a tunable security/overhead parameter. If routing leaks, signs are correlated, dummy rotations are distinguishable, grid labels collide, optimizer-generated angles are correlated, or cross-shot grouping is possible, the ambiguity can degrade.
 \item The trap-based detection layer presented here detects broad classes of deviations that affect the verifier circuit, but it is an embedded statistical detection layer rather than a complete malicious-security proof. Full malicious security would require a trap/authentication framework with its own soundness proof, or a simulator-based verifiable-delegation construction.
 \item Protocol \ref{Protocol: 4} requires the non-total-collusion model, hidden matching, side-channel control, the explicit \(\epsilon_{\mathrm{key}}\) key-hiding condition, and, in the recommended high-security variant, the common-node Pauli refresh of Assumption \ref{ass:protocol4-coalitions}. The computational servers may pool value-share transcripts, but the privacy claims fail or degrade if the full coalition $\{S_1,S_2,R\}$ forms, if the common-node matching or refresh share leaks, or if structural side channels reveal grouped angle shares or aligned Pauli keys. Collusion of $S_1$ and $S_2$ alone is not automatically fatal only when the hidden matching compiler preserves the $\epsilon_{\mathrm{key}}$-key-hiding condition, or when the common-node Pauli refresh supplies an additional hidden uniform share. If the matching is trivial, low-entropy, or leaked through side channels, then the aligned Pauli frame may be reconstructed and the Protocol \ref{Protocol: 4} state-privacy claim degrades accordingly.
 \item In the zero-qubit setting, the client does not obtain the same direct state-level privacy guarantee as in the client-side QOTP protocols because it never directly touches the quantum register; Protocol \ref{Protocol: 4} instead obtains state privacy under persistent split-QOTP plus \(\epsilon_{\mathrm{key}}\) key hiding. The common node and any measurement node are part of this trust and leakage model. Its shuffled $r$-share sign-randomized angle sharing gives transcript-level angle unlinkability, not malicious security or universal blindness by itself. Stronger malicious or verifiable guarantees require quantum authentication, trap-based VBQC, UBQC/VBQC-style constructions~\cite{Universal_Blind,FitzsimonsKashefi2017VBQC}, or equivalent mechanisms with their own proofs.
 \item Authenticated QOTP is an optional high-resource strengthening, summarized in Appendix \ref{app:authenticated-qotp}; it changes the encoded data blocks and overhead assumptions rather than upgrading the base protocols for free.
\end{itemize}

\subsection{\texorpdfstring{Failure modes when assumptions are violated.}{Failure modes when assumptions are violated.}}
\begin{enumerate}
 \item \textbf{Degenerate matching.} If \(m=1\) in Protocol \ref{Protocol: 4} and no common-node Pauli refresh is used, then \(\sigma_R\) is trivial. A coalition \(\{S_1,S_2\}\) pooling both share tables can reconstruct the aligned Pauli key. Therefore the key-hiding condition fails unless an additional hidden refresh share is used.
 \item \textbf{Leaked grouping in angle sharing.} If the server can group all \(r\) visible shares of a private rotation and infer the hidden signs, it can reconstruct the signed sum and recover the finite-grid angle label. If signs remain hidden but grouping is known, ambiguity is reduced to \(\Theta_p(\alpha)\), which may be small for small \(r\), repeated angles, or structured optimizer-generated parameters.
 \item \textbf{Visible zero dummies.} If dummy or idle rotations are sent as recognizable zero rotations, the server can distinguish absent edges or idle slots. This motivates masked identities \(R_A(-\eta)R_A(\eta)\) compiled, shuffled and routed like real rotations.
 \item \textbf{Cross-shot alignment.} Even with fresh QOTP keys and fresh angle masks, if the same logical edge or parameter can be aligned across shots through timing, batch labels, routing latency or optimizer traces, the server may intersect candidate angle sets across repetitions and reduce ambiguity.
\end{enumerate}

\section{Conclusions}
This work presents several protocols and techniques for delegated quantum computation with privacy guarantees adapted to the client's resources and to the need to protect both data and quantum operations. By separating QOTP state privacy under declared leakage from compiler- and leakage-function-dependent structural hiding, matching-hidden split-QOTP and output privacy and trap-based detection, we obtain a family of protocols whose guarantees can be stated explicitly for different client capabilities. Fully capable $M$-qubit clients obtain QOTP-based state privacy while delegating public Clifford computation and keeping private or non-Clifford operations local when they fit in memory. Clients with independent single-qubit devices require compilation into public Clifford layers and local single-qubit layers, plus routing assumptions for structural hiding. Restricted devices can use finite-grid routing-hidden sign-randomized $r$-share Pauli-axis rotation ambiguity for selected private one-qubit rotations, but only under the hidden-sign, hidden-grouping, finite-grid and side-channel assumptions stated above. Classical-client variants form a separate non-total-collusion branch that replaces client-side QOTP control by a multi-party persistent matching-hidden split-QOTP maintained through computational servers and the common node, together with shuffled $r$-share sign-randomized angle sharing and leakage-conditioned split-frame readout. In this branch, the state claim is state privacy under persistent split-QOTP plus \(\epsilon_{\mathrm{key}}\) key hiding, the angle claim is transcript-level angle unlinkability, and the construction is not universal blindness or malicious security without separate verification or authentication. Across the hierarchy, traps add statistical detection of certain deviations but not a complete malicious-security proof. Future work could reduce the number of qubit transmissions while maintaining the stated privacy levels, since these transmissions increase circuit execution time and carry an operational cost. Another direction is to strengthen the non-Clifford and detection components so that the high-level protocols presented here can be upgraded to stronger end-to-end security notions, ideally in a fault-tolerant setting.

\section*{Acknowledgments}
This work was developed within the activities of the Quipucamayocs group of the QuantumQuipu community.

\section*{Declaration of generative AI and AI-assisted technologies}

During the preparation and revision of this manuscript, the authors used ChatGPT, specifically GPT-5.5 Pro, to assist with critical review, identification of possible inconsistencies or weaknesses, and limited improvements to clarity and language.

The authors did not use the tool to generate the original algorithms, protocols, security assumptions, proofs, technical claims, or scientific conclusions. All AI-assisted suggestions were reviewed, verified, and edited by the authors, who take full responsibility for the final content of the manuscript.

\bibliographystyle{unsrt}
\bibliography{references}

\newpage
\appendix
\section{\texorpdfstring{Authenticated QOTP high-resource variant}{Authenticated QOTP high-resource variant}}\label{app:authenticated-qotp}
The trap-based layer of Sec. \ref{sec: verification} and authenticated QOTP address related but different aspects of malicious deviations. The trap-based layer is an external statistical detection layer: it hides verifier locations among computational locations and detects deviations when the attack affects a verifier location and changes its expected classical outcome. By contrast, an authenticated QOTP layer would attach integrity checks directly to the delegated quantum data blocks. Thus, the two mechanisms are complementary rather than identical. In the terminology used above, the trap-based detection layer is the baseline check for this paper's leakage-dependent protocols, whereas authenticated QOTP is an optional strengthening that changes the encoded form of the delegated data.

The QOTP layer used in Protocols \ref{Protocol: 1}--\ref{Protocol: 3} gives information-theoretic state privacy, but bare QOTP does not give integrity. A malicious server may apply an arbitrary quantum operation to an encrypted register, and the client may receive a corrupted logical state without detecting it. A possible strengthening, available only for clients with sufficient quantum memory and encoding capability, is to replace the bare QOTP encryption
\begin{equation}
 \rho\mapsto P_{x,z}\rho P_{x,z}^{\dagger}
\end{equation}
by an authenticated encryption map
\begin{equation}
 \rho\mapsto \operatorname{AuthEnc}_k(\rho),
\end{equation}
where $k$ is a secret classical authentication key. The corresponding decoding map satisfies, informally,
\begin{equation}
 \operatorname{AuthDec}_k\!\left(\mathcal{A}\!\left(\operatorname{AuthEnc}_k(\rho)\right)\right)
 \in\{\rho,\bot\}
\end{equation}
except with probability at most $\varepsilon_{\mathrm{auth}}(s)$, where $\mathcal{A}$ is the server's attack map, $s$ is a security parameter and $\bot$ denotes rejection. Equivalently, for one authenticated check,
\begin{equation}
 \Pr[\mathsf{accept}\wedge\mathsf{corrupted}]
 \leq \varepsilon_{\mathrm{auth}}(s).
\end{equation}
For $T$ authenticated transmissions or decoding checks, the basic union-bound estimate gives
\begin{equation}
 \Pr[\mathsf{accept}\wedge\mathsf{corrupted}]
 \leq T\varepsilon_{\mathrm{auth}}(s).
\end{equation}
This is only a statement of the intended authentication failure parameter for the authenticated layer; it is not, by itself, a complete composable malicious-security proof for the delegated computation.

The notation $\operatorname{AuthEnc}_k$ should not be understood as a new primitive introduced in this paper, but as abstract notation for a quantum authentication scheme. Quantum authentication schemes were introduced to authenticate quantum messages using a shared classical secret key: they encode an $m$-qubit message into a larger physical block, encrypt the block and later allow the receiver either to recover the original logical state or to reject \cite{Barnum2002Authentication}. One representative trap-code-style picture, useful for interpreting the notation but not claimed to be implemented by Protocols \ref{Protocol: 1}--\ref{Protocol: 4}, is
\begin{equation}
\begin{aligned}
 \operatorname{AuthEnc}_k(\rho)
 &=
 P_{a,b}\Pi_{\pi}E_c
 \left(
 \rho
 \otimes |0\rangle\langle 0|^{\otimes t_Z}
 \otimes |+\rangle\langle +|^{\otimes t_X}
 \right)
 E_c^{\dagger}\Pi_{\pi}^{\dagger}P_{a,b}^{\dagger},
\end{aligned}
\end{equation}
with
\begin{equation}
 k=(c,\pi,a,b).
\end{equation}
Here $c$ specifies the authentication code or encoding choice, $\pi$ specifies a secret permutation of the physical positions, and $a,b$ are the Pauli keys of the QOTP applied to the full physical block. The map $E_c$ encodes the logical register together with auxiliary trap or syndrome registers into a larger physical block. The $|0\rangle$ traps are sensitive to bit-flip-type disturbances, while the $|+\rangle$ traps are sensitive to phase-flip-type disturbances. The secret permutation $\Pi_{\pi}$ hides from the server which physical positions correspond to data, traps or checks, and the physical QOTP $P_{a,b}$ hides the full authenticated block from the server. When the block is returned, the client applies the inverse Pauli mask, undoes the permutation, decodes using $E_c^{\dagger}$ or the corresponding decoding map, and checks the trap or syndrome registers. If all checks pass, the client accepts the decoded logical register; if any check fails, the client outputs $\bot$ and aborts. This expression is only a representative trap-code-style authentication picture, related to known constructions for authenticated quantum data \cite{Broadbent2013QuantumOneTime}; other quantum authentication schemes may use different codes, traps, syndromes or key structures.

This strengthening is not resource-free. The bare QOTP has essentially one physical qubit per logical qubit, apart from the classical keys. By contrast, authenticated QOTP requires extra physical qubits for traps, syndrome registers or authentication redundancy. In the authentication scheme of Barnum et al., an $m$-qubit quantum message is encoded into $m+s$ qubits, with authentication error decreasing exponentially in the security parameter $s$ \cite{Barnum2002Authentication}. Thus, if an authentication code maps an $m$-qubit logical register to an $(m+s)$-qubit authenticated block, the qubit overhead factor is
\begin{equation}
 \frac{m+s}{m}
 =
 1+\frac{s}{m}.
\end{equation}
The client's effective logical memory is correspondingly reduced. Ignoring additional workspace, a client with $M$ physical qubits can store roughly
\begin{equation}
 M_{\mathrm{logical}}\approx m
 \left\lfloor\frac{M}{m+s}\right\rfloor
\end{equation}
logical qubits when using blocks of size $m$. In the qubit-by-qubit illustrative case where each logical qubit is protected separately with two additional trap qubits, one obtains a three-qubit physical block per logical qubit and therefore
\begin{equation}
 M_{\mathrm{logical}}=
 \left\lfloor\frac{M}{3}\right\rfloor.
\end{equation}
This qubit-by-qubit picture is illustrative and is not necessarily an optimal authentication code. Additional workspace for encoding, decoding, syndrome extraction or trap checks can reduce this number further.

For this reason, authenticated QOTP should be regarded as a high-resource, high-overhead variant mainly of Protocol \ref{Protocol: 1}, not as the default mechanism for the low-resource protocols. In Protocol \ref{Protocol: 2}, full block-level authentication would require either multi-qubit encoding and decoding capabilities or an additional trusted local module capable of preparing and checking authenticated blocks. If the client only has independent single-qubit devices, then full quantum authentication is not obtained from the existing assumptions. In Protocol \ref{Protocol: 4}, the client is classical and therefore cannot directly prepare, encrypt, decode or verify authenticated quantum states. In that setting, authenticated QOTP would require additional primitives or trusted or non-colluding quantum parties and is outside the basic protocol.

Delegated public Clifford operations on authenticated data would also need authenticated logical implementations. That is, for a public Clifford $C$, the protocol would need a physical operation $\widetilde{C}$ and a key update $k\mapsto k'$ such that
\begin{equation}
 \widetilde{C}\operatorname{AuthEnc}_k(\rho)\widetilde{C}^{\dagger}
 =
 \operatorname{AuthEnc}_{k'}(C\rho C^{\dagger}).
\end{equation}
If such an authenticated logical implementation is not supplied, the register must be returned to the client, decoded and checked, operated on locally, and freshly authenticated before being delegated again. Non-Clifford operations still require the additional mechanisms already discussed in the paper; authentication does not remove the need for adapted non-Clifford gadgets, local client operations or another explicit encrypted-evaluation primitive.

Therefore, authenticated QOTP is a natural optional strengthening for clients with enough quantum memory and coding capability, because it upgrades delegated storage from privacy-only to privacy-plus-integrity. However, it introduces precisely the kind of qubit and operation overhead that this work tries to avoid in user-level and low-resource settings. The trap-based detection layer of Sec. \ref{sec: verification} is therefore kept as the baseline verification mechanism, while authenticated QOTP is mentioned only as a high-resource alternative or complement.

\section{\texorpdfstring{Optional non-Clifford gadget route for Protocols 2 and 3}{Optional non-Clifford gadget route for Protocols 2 and 3}}\label{app:optional-nonclifford-gadgets}
This appendix records an optional extension path, not a mechanism used in the base protocols. In delegated encrypted quantum computation, Clifford gates are comparatively natural because their action on QOTP-encrypted data can be absorbed into Pauli-frame updates. Non-Clifford gates, such as $T$ or other $\pi/8$-type rotations, usually require extra interaction, magic states, gadgets, or more involved key-evaluation machinery.

Broadbent's delegated private quantum computation protocol is one representative example: Clifford operations are handled more directly, while the $\pi/8$ gate requires the client to prepare and send an additional random auxiliary qubit together with classical communication \cite{Delegating_private}. Broadbent and Jeffery's quantum homomorphic encryption construction for circuits of low $T$-gate complexity gives another route, with overhead that grows with the non-Clifford part of the circuit \cite{T_gate}.

These techniques can be viewed as possible upgrades for Protocol \ref{Protocol: 2}, and for Protocol \ref{Protocol: 3} when the restricted client is supplemented with enough quantum capability to prepare the required auxiliary states or participate in the required interaction. They would change the resource assumptions, leakage analysis, and overhead, and therefore are not part of the base protocol statements.

This appendix does not extend Protocol \ref{Protocol: 4}. In Protocol \ref{Protocol: 4} the client is assumed not to have the non-Clifford quantum capability required by these gadgets; therefore these techniques are mentioned only as an optional upgrade path for Protocols \ref{Protocol: 2} and \ref{Protocol: 3}.

\end{document}